\documentclass[twocolumn,english,reprint, superscriptaddress, breaklinks=true, showkeys, showpacs=false, nofootinbib]{revtex4-2}

\usepackage[T1]{fontenc}
\usepackage{color}
\usepackage{babel}
\usepackage{float}
\usepackage{amssymb}
\usepackage{listings}
\usepackage{graphicx}
\usepackage{amsmath}
\usepackage{svg}
\usepackage{booktabs} 
\usepackage[normalem]{ulem}

\usepackage{subcaption}

\usepackage{listings}
\usepackage{xcolor} 
\definecolor{LightGray}{gray}{0.9}
\definecolor{DarkGreen}{rgb}{0,0.5,0}

\usepackage{array}
\usepackage{ragged2e}

\usepackage{bbding}
\usepackage{amssymb}
\usepackage{xcolor}
\newcommand{\greencheck}{{\color{green}\CheckmarkBold}}
\newcommand{\redcross}{{\color{red}\XSolidBold}}

\usepackage[unicode=true,pdfusetitle, bookmarks=true,bookmarksnumbered=false,bookmarksopen=false, breaklinks=true,pdfborder={0 0 0},backref=false,colorlinks=true]{hyperref}
\usepackage{breakurl}
\definecolor{Code}{rgb}{0,0,0}
\definecolor{Decorators}{rgb}{0.5,0.5,0.5}
\definecolor{Numbers}{rgb}{0.5,0,0}
\definecolor{MatchingBrackets}{rgb}{0.25,0.5,0.5}
\definecolor{Keywords}{rgb}{0,0,1}
\definecolor{self}{rgb}{0,0,0}
\definecolor{Strings}{rgb}{0,0.63,0}
\definecolor{Comments}{rgb}{0,0.63,1}
\definecolor{Backquotes}{rgb}{0,0,0}
\definecolor{Classname}{rgb}{0,0,0}
\definecolor{FunctionName}{rgb}{0,0,0}
\definecolor{Operators}{rgb}{0,0,0}
\definecolor{Background}{rgb}{0.98,0.98,0.98}
\lstdefinelanguage{Python}{
numbers=left,
numberstyle=\footnotesize,
numbersep=1em,
xleftmargin=1em,
framextopmargin=2em,
framexbottommargin=2em,
showspaces=false,
showtabs=false,
showstringspaces=false,
frame=l,
tabsize=4,
basicstyle=\ttfamily\small\setstretch{1},
backgroundcolor=\color{Background},
commentstyle=\color{Comments}\slshape,
stringstyle=\color{Strings},
morecomment=[s][\color{Strings}]{"""}{"""},
morecomment=[s][\color{Strings}]{'''}{'''},
morekeywords={import,from,class,def,for,while,if,is,in,elif,else,not,and,or,print,break,continue,return,True,False,None,access,as,,del,except,exec,finally,global,import,lambda,pass,print,raise,try,assert},
keywordstyle={\color{Keywords}\bfseries},
morekeywords={[2]@invariant,pylab,numpy,np,scipy},
keywordstyle={[2]\color{Decorators}\slshape},
emph={self},
emphstyle={\color{self}\slshape},
}

\makeatletter
\@ifundefined{textcolor}{}
{%
 \definecolor{BLACK}{gray}{0}
 \definecolor{WHITE}{gray}{1}
 \definecolor{RED}{rgb}{1,0,0}
 \definecolor{GREEN}{rgb}{0,1,0}
 \definecolor{BLUE}{rgb}{0,0,1}
 \definecolor{CYAN}{cmyk}{1,0,0,0}
 \definecolor{MAGENTA}{cmyk}{0,1,0,0}
 \definecolor{YELLOW}{cmyk}{0,0,1,0}
}

\hypersetup{colorlinks=true,citecolor=blue,linkcolor=cyan,urlcolor=blue,filecolor= green, breaklinks=true}
\usepackage{url}
\usepackage{breakurl}

\makeatother

\begin{document}

\title{QuForge: A Library for Qudits Simulation}

\author{Tiago de Souza Farias\href{https://orcid.org/0000-0002-6697-9333}{\includegraphics[scale=0.05]{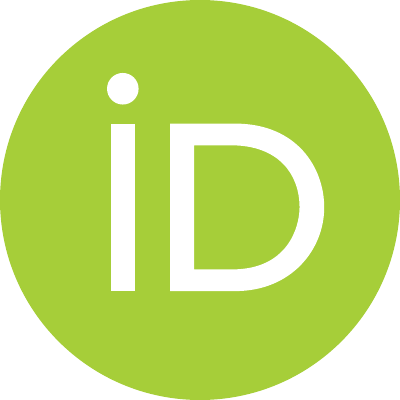}}}
\email{tiago.farias@ufscar.br}
\affiliation{Physics Department, Center for Exact Sciences and Technology, Federal University of S\~ao Carlos, 13565-905, S\~ao Carlos, SP, Brazil}

\author{Lucas Friedrich\href{https://orcid.org/0000-0002-3488-8808}{\includegraphics[scale=0.05]{orcidid.pdf}}}
\email{lucas.friedrich@acad.ufsm.br}
\affiliation{Physics Department, Center for Natural and Exact Sciences, 
Federal University of Santa Maria, 97105-900,
Santa Maria, RS, Brazil}

\author{Jonas Maziero\href{https://orcid.org/0000-0002-2872-986X}{\includegraphics[scale=0.05]{orcidid.pdf}}}
\email{jonas.maziero@ufsm.br}
\affiliation{Physics Department, Center for Natural and Exact Sciences,
Federal University of Santa Maria, 97105-900,
Santa Maria, RS, Brazil}

\begin{abstract}
Quantum computing with qudits, an extension of qubits to multiple levels, offers promising advantages in information representation and computational density. Despite its potential, tools for qudit-based quantum computation remain underdeveloped. This article presents QuForge, a Python-based library that introduces a novel computational framework for simulating quantum circuits with arbitrary qudit dimensions. QuForge distinguishes itself by providing scalable and extensible features, such as a complete set of customizable quantum gates, support for sparse matrix representations, and compatibility with GPU and TPU accelerators to optimize performance. By leveraging differentiable frameworks, QuForge accelerates simulations and facilitates quantum machine learning and algorithm development research. Through empirical demonstrations and benchmarks, we highlight the capabilities of the library to address scalability challenges and enable advances in quantum information science, establishing it as a potential tool for advancing research and applications in high-dimensional quantum computing.
\end{abstract}


\maketitle

\section{Introduction}

Differentiable programming frameworks such as PyTorch \cite{paszke2019pytorch} and JAX \cite{jax2018github} have revolutionized the development of computational tools by offering robust infrastructures for constructing algorithms composed of differentiable components \cite{Blondel2024}. This inherent differentiability enables data-driven optimization of program parameters, a cornerstone of advances in machine learning and deep learning \cite{Goodfellow2016}.

These frameworks are designed with versatility, supporting seamless execution across various hardware platforms, from CPUs to GPUs and TPUs \cite{tpupaper}. This adaptability not only accelerates execution but also extends the applicability of differentiable programming to a wide array of scientific challenges.

In quantum computing, the primary focus has traditionally been on qubits \cite{chae_elementary_2024}, representing two-level quantum systems capable of existing in a superposition of states 0 and 1. Quantum algorithms exploit these properties, along with entanglement, to achieve computational advantages. However, qudits extend this paradigm by allowing superpositions of more than two states, offering denser information encoding and expanded computational capabilities \cite{Brylinski2001UniversalQG, Wach_2023}.

Qudits present unique advantages over qubits by enhancing efficiency and scalability, allowing the resolution of complex problems with fewer quantum resources \cite{goyal_qudit-teleportation_2014, qudit_info2}. Furthermore, expanded state space facilitates the exploration of new quantum phenomena, advancing fields such as quantum simulation \cite{quditsimulation, tacchino_proposal_2021}, cryptography \cite{qudit_crypt}, and communication \cite{quditcomm}. The potential for more robust error correction schemes and quantum gate operations further underscores their importance \cite{chizzini_quantum_2022, qudit_error2}.

Quantum machine learning (QML) has emerged as an important subfield within quantum information science, motivated by the prospect of surpassing classical machine-learning paradigms in both data efficiency and computational throughput. Several studies have demonstrated that quantum models can achieve comparable predictive accuracy with fewer training samples and may offer algorithmic speed-ups in optimization loops and inference routines \cite{caro_generalization_2022, ding2024quantumactivelearning, liu_rigorous_2021, liao_quadratic_2024}. In addition, QML architectures have shown particular promise in domains characterized by high-dimensional feature spaces and complex energy landscapes, most notably in quantum chemistry and material discovery \cite{cerezo_challenges_2022, grana_materials_2025, sajjan_quantum_2022}. 

While the majority of QML research has focused on qubits, leveraging higher-dimensional quantum systems could further enhance representational capacity. Consequently, incorporating qudit degrees of freedom into QML frameworks holds potential to accelerate learning and expand applicability to computationally demanding scientific problems.

Despite these advantages, the lack of accessible and scalable tools for qudit-based quantum computation has hindered progress. Addressing this gap requires a high-level library that supports qudit simulation and leverages modern computational paradigms to enhance scalability and flexibility.

Here, we introduce QuForge, a Python-based library designed to simulate qudit-based quantum circuits. QuForge integrates differentiable programming principles and supports execution on various hardware platforms. A central innovation is its use of sparse matrix operations \cite{Scott2023}, significantly reducing memory and computational overhead for large qudit systems. By enabling broad access to high-dimensional quantum computation, QuForge provides researchers with the tools necessary to systematically investigate the uncharted capabilities of qudits, fostering significant advancements in quantum technologies.

The sequence of this article is organized as follows. In Sec. \ref{sec:other_work} we discuss some related works from the literature. In Sec. \ref{sec:quforge}, we present the QuForge library, discussing the quantum gates used (Sec. \ref{sec:qgates}) and theirs sparsity (Sec. \ref{sec:sparsity}), and also the code style (Sec. \ref{sec:codestyle}). In Sec. \ref{sec:examples}, we present examples of results obtained using the QuForge library applied to the Deutsch-Jozsa, Grover, and variational quantum algorithms. Finally, we give our conclusions in Sec. \ref{sec:conc}. The matrix form of the quantum gates and their sparse representation are discussed in the Appendix \ref{appendix:1}.

\section{Related work}
\label{sec:other_work}

The study of qudit quantum computation dates back to the early 21st century, as initially explored by Jean-Luc Brylinski and Ranee Brylinski in their work on universal quantum gates \cite{Brylinski2001UniversalQG}. Since then, extensive research has been conducted on multi-level quantum bits. This research encompasses various aspects of quantum computing, demonstrating the potential and versatility of qudits in advancing quantum algorithms and applications.

For instance, the Quantum Approximate Optimization Algorithm (QAOA), a prominent quantum machine learning algorithm, has been successfully generalized to incorporate qudits, thereby expanding its applicability and performance in solving combinatorial optimization problems \cite{Deller_2023}. Similarly, quantum variational algorithms, which are pivotal in numerous quantum computing tasks \cite{alchieri_introduction_2021, cerezo_variational_2021}, have also been adapted to leverage the advantages offered by qudits \cite{rocajerat2023qudit, qudit_ml2}. These advancements underscore the significant role qudits can play in enhancing the efficiency and effectiveness of quantum algorithms.

Beyond computational applications, developing robust error correction techniques for qudits is critical to realizing practical and scalable quantum computing systems. Recent studies have addressed this need by proposing novel error correction codes and strategies specifically designed for multi-level quantum systems \cite{Fischer_2023}. These efforts are essential in mitigating the effects of decoherence and operational errors, thus paving the way for more reliable and fault-tolerant quantum qudit devices.

CUDA \cite{cudapaper}, a parallel computing platform and programming model developed by NVIDIA, facilitates the execution of operations on Graphics Processing Units (GPUs). These GPUs are equipped with thousands of cores capable of performing numerous elementary operations simultaneously, thereby significantly enhancing the speed of algebraic computations through parallel processing capabilities in contrast to traditional Central Processing Units (CPUs). 

Given that the evolution of quantum circuits inherently involves complex matrix multiplications, GPUs are exceptionally well-suited for addressing such computational challenges. This advantage is particularly relevant in qudit simulations, where efficiently handling high-dimensional quantum systems requires substantial computational resources.

Tensor Processing Units (TPUs) \cite{tpupaper}, designed by Google, represent a specialized hardware acceleration aimed explicitly at speeding up matrix operations, which are pivotal in deep learning model training and inference. These fundamentally matrix-based operations underpin both deep learning and quantum computing methodologies. The architecture of TPUs is optimized for high-throughput, low-precision arithmetic operations, a requirement that aligns closely with the computational demands of deep neural networks. Given the analogous nature of the underlying computations, quantum computing simulations stand to gain substantially from the deployment of TPUs. The parallel processing capabilities and optimized matrix computation functionalities of TPUs offer a promising direction for enhancing the efficiency of simulations of quantum circuits, particularly in the domain of qudit-based quantum computing.

PyTorch, a comprehensive open-source machine learning library, facilitates the seamless transition between computational devices, namely, CPUs, GPUs, and TPUs, allowing for flexible computational resource allocation. Furthermore, the framework incorporates fundamental differentiable building blocks, enabling the construction of differentiable graphs. This capability is critical for optimizing parameterized operations via gradient-based optimization methods. The convergence of these features, PyTorch's device-agnostic computation and its robust support for differentiable programming, alongside the shared algebraic foundation between quantum computing and deep learning, positions PyTorch as an excellent platform for developing a qudit simulation library.

In the landscape of quantum software, it is important to differentiate between platforms designed primarily for algorithm expression and those intended for circuit simulation.

Quantum frameworks such as Qiskit \cite{qiskit2024}, Pennylane \cite{bergholm2022pennylane}, and Cirq \cite{developers_cirq_2024} provide comprehensive environments for designing and implementing quantum algorithms. However, their support for qudit-based computation is typically limited. For example, Qiskit and Pennylane usually support qudits only up to qutrits and often require manual construction of qudit gates or special workarounds to simulate higher-dimensional operations. While these frameworks excel in user-friendly APIs and integration with quantum hardware, they generally do not offer an extensive set of pre-built qudit-based quantum gates.


On the simulation side, several libraries have emerged to facilitate the classical simulation of quantum circuits. Standalone or integrated simulators, such as quDiet \cite{chatterjee_qudiet:_2023} and Jet \cite{vincent_jet:_2022}, support the simulation of qudit circuits and incorporate advanced features like GPU acceleration and sparse matrix representations. These tools are optimized for efficiently handling the high-dimensional Hilbert spaces inherent to qudits. However, many of these simulators lack a high-level interface for algorithm expression, often requiring users to manually implement quantum gates or construct circuits with a lower level of abstraction.

Additional libraries in this domain include sQUlearn \cite{kreplin2024squlearn}, which targets quantum machine learning and supports implementations on the same hardware as Qiskit and Pennylane; Google's Circ library \cite{developers_cirq_2024}, which enables qudit simulations but necessitates manual gate implementation; and torchQuantum \cite{torchquantum}, a PyTorch-based library that, despite its ML-friendly design, currently lacks native qudit support. MQT Qudits \cite{mqtqudits}, on the other hand, offers multidimensional qudit support but does not integrate optimization features.

In contrast, our proposed library aims to fill these gaps by providing comprehensive support for qudit simulation with ready-to-use quantum gates and multidimensional support, while also user-friendly for quantum machine learning applications. This unique combination of features distinguishes our library from existing ones and addresses the current limitations in the field of qudit simulation.

Table \ref{table:1} provides a comprehensive summary of the related libraries, highlighting their respective capabilities and features. The table categorizes these libraries based on several key criteria: support for qudits, GPU simulation capabilities, the availability of pre-implemented qudit gates, their suitability for quantum machine learning applications, their ability to work with sparse operations, and support for integrating qudits with different dimensions in the same circuit.

\begin{widetext}

\begin{table}[h!]
\setlength{\tabcolsep}{2pt}
\begin{center}
\begin{tabular}{ | c |c | c | c | c | c | c |} 
  \hline
  Library & Qudit support & GPU support & Qudit gates & QML friendly & Sparse support & Multidimensional support\\
  \hline
  Qiskit \cite{qiskit2024} & up to qutrit & \greencheck & \redcross & \redcross & Pauli Gates & \redcross\\ 
  \hline
  Pennylane \cite{bergholm2022pennylane} & up to qutrit & \greencheck & \redcross & \greencheck & \greencheck & \redcross \\
  \hline
  sQUlearn \cite{kreplin2024squlearn} & \redcross & \redcross & \redcross & \greencheck & Pauli Gates & \redcross \\ 
  \hline
  quDiet \cite{chatterjee_qudiet:_2023} & \greencheck & \greencheck & \greencheck & \redcross & \greencheck & \greencheck \\
  \hline
  circ \cite{developers_cirq_2024} & \greencheck & \greencheck & \redcross & \redcross & \greencheck & \redcross \\ 
  \hline
  Jet \cite{vincent_jet:_2022} & \greencheck & \greencheck & \redcross & \greencheck & \redcross & \redcross \\
  \hline
  torchQuantum \cite{torchquantum} & \redcross & \greencheck & \redcross & \greencheck & \redcross & \redcross\\
  \hline
  MQT Qudits \cite{mqtqudits} & \greencheck & \redcross & \greencheck & \redcross & \greencheck & \greencheck \\ 
  \hline
  \hline
  \textbf{QuForge} & \greencheck & \greencheck & \greencheck & \greencheck & \greencheck & \greencheck\\ 
  \hline
\end{tabular}
\caption{Comparison of quantum simulation libraries based on key features relevant to quantum circuits: qudit support, GPU compatibility, availability of pre-implemented qudit gates, ease of integration with quantum machine learning, support for sparse operations, and multidimensional support.}
\label{table:1}
\end{center}
\end{table}

\end{widetext}

\section{QuForge}
\label{sec:quforge}

To develop a library for qudit simulation based on a circuit-based architecture, it is important to establish a comprehensive set of quantum logical gates that can operate effectively at the qudit level. These gates are primarily generalizations of qubit gates, adapted for multi-level quantum systems. The generalization process involves extending the operational principles of qubit gates to accommodate the additional states available in qudits. These quantum gates can be represented in matrix form, enabling their application to the state vector representation of the wave function. This matrix representation is crucial for performing quantum operations and simulations, as it provides a mathematical framework for manipulating the states of qudits.

To create a user-friendly high-level library, it is essential to devise methods for automatically generating these matrix representations for any given dimension of the qudit. This involves developing algorithms that can compute the necessary matrix elements based on the specific properties and requirements of the qudit system. Such algorithms must be robust and versatile, handling various dimensions and ensuring accurate quantum operations. Furthermore, incorporating these automated matrix generation techniques into the library facilitates integration and usability for researchers and practitioners, as it allows users to quickly define and manipulate qudits without needing to manually construct the underlying matrices, thereby streamlining the process of qudit simulation and broadening the accessibility of advanced quantum computing tools.

\subsection{Quantum Gates}
\label{sec:qgates}

QuForge has implemented the essential quantum gates required for achieving universal quantum computation, meaning that any quantum algorithm or operation can be decomposed into a set of these fundamental gates. Although multiple sets of gates qualify as universal, including additional complementary gates can significantly simplify the abstraction and implementation of algorithms in quantum circuits. 

Therefore, the library provides a broader set of gates beyond the minimal universal set to enhance usability and flexibility. The primary gates implemented in QuForge include the \texttt{Hadamard} gate, which, in the context of qudits, is also called the Fourier gate and is crucial for creating superposition states. The controlled-NOT (\texttt{CNOT}) gate, which facilitates entanglement between qudits, is another fundamental gate the library provides.

Additionally, rotation gates, which are particularly important for quantum machine learning algorithms, are included. These gates feature angle parameters that can be optimized during the training of quantum models, making them indispensable for parameterized quantum circuits. 

The \texttt{SWAP} gate is also implemented in QuForge, which is vital for specific quantum algorithms that require qubit or qudit state exchanges. This gate is essential for reordering quantum states in a circuit, which can be necessary for efficient algorithm implementation and optimization. By incorporating these key gates and more, QuForge ensures that users have a comprehensive toolkit for developing and simulating a wide range of quantum algorithms and applications.

Further implemented gates in QuForge include the generalized Gell-Mann gates, \texttt{Toffoli} gate, controlled rotation gates, and the multi-controlled NOT gate. The Gell-Mann matrices extend the concept of Pauli matrices to higher dimensions, providing a tool for operations in multi-level quantum systems. The \texttt{Toffoli} gate, or controlled-controlled NOT gate, is a fundamental gate for implementing reversible computation and error correction. Controlled rotation gates can be used for fine-tuning quantum states and are particularly useful for quantum machine learning. The multi-controlled NOT gate extends the entangling capabilities of the \texttt{CNOT} gate to more complex systems with multiple control qudits.

Additionally, QuForge allows users to create custom gates by inputting the associated matrix. This feature provides flexibility and adaptability, enabling users to implement and experiment with novel quantum gates tailored to their research needs and applications.

Appendix \ref{appendix:1} provides a detailed listing of the matrix forms of the quantum gates implemented in QuForge. Most of these gates are generalizations of qubit gates, meaning that for qudits with dimension $D=2$, their structure will mirror that of standard qubit gates. The primary difference lies in the size and elements of the matrices, which depend on the dimension of the qudit. As the dimension increases, the matrices expand accordingly to accommodate the additional states of the qudit system. 

\subsection{Library Architecture}

In QuForge, we have adopted a modular and extensible architecture built around several core components. At the heart of the library is the Circuit class, which serves as a container for constructing quantum circuits. This class manages the sequential application of quantum operations and gates on quantum states. The State class represents the quantum state of the system, encapsulating the complex amplitude vector for qudits. These classes enable users to define, simulate, and measure quantum circuits.

The library provides a comprehensive suite of pre-implemented quantum gate classes, such as the generalized \texttt{Hadamard/Fourier} gate, \texttt{CNOT}, and rotation gates (\texttt{RX, RY, RZ}). These gate classes encapsulate the mathematical operations, with their matrix representations, necessary for manipulating qudit states. Additionally, QuForge supports custom gate definitions, allowing users to input matrix representations directly when a novel operation is required.

To accommodate different programming preferences, QuForge offers multiple interfaces:

\begin{itemize}
    \item Algorithm-Centric Style: Inspired by traditional quantum circuit libraries (like Qiskit), this style allows users to add gates to the circuit sequentially.

    \item Machine Learning-Centric Style: Inspired by frameworks such as PyTorch, this approach represents circuits as differentiable modules, facilitating gradient-based optimization in quantum machine learning applications.

    \item Designed for data-centric applications, this style integrates classical pre-processing with quantum circuit execution.
\end{itemize}

These components are designed to interact in a cohesive workflow: a user begins by defining a quantum circuit using the Circuit class, selects and applies a series of quantum gates (either pre-defined or custom) to build the circuit, initializes a quantum state with the State class, and then executes the circuit to obtain a final state. Measurement functions then provide the output in the form of probabilities or histograms.

For a more holistic view, Figure \ref{fig:diagram}, presented in the appendix \ref{appendix:1}, illustrates a UML-like diagram showing the relationships between the main classes (\texttt{Circuit}, \texttt{State}, \texttt{Gate} classes, \texttt{Module}, and \texttt{Measurement} functions) and how they work together to simulate qudit-based quantum computation. This design streamlines the simulation process and can leverage hardware acceleration, making the library scalable for larger, more complex systems.

\subsection{Sparsity to lower memory consumption}
\label{sec:sparsity}

Certain quantum gates exhibit a relatively low count of non-zero elements that scales with the qudit dimensionality. This sparsity can be exploited by directly constructing sparse matrix representations, thereby reducing the memory footprint required to store gate operations. It is important to note that, while the overall state vector is represented as a dense vector with $d^n$ elements (where $d$ is the qudit dimension and $n$ is the number of qudits), the memory cost associated with the gate representations can be much higher. In particular, a dense matrix representation for a single gate acting on $n$ qudits has $d^{2n}$ elements. When using gradient-based optimization, every gate in the circuit must be stored and differentiated during the simulation, meaning that the cumulative memory for storing these dense matrices can dominate, even though the state vector itself is smaller.

Sparse representations, which focus exclusively on storing the indices and values of non-zero elements, offer a significant memory saving. For example, consider the generalized Gell-Mann symmetric gate $S_x^{j,k}$ defined as: 
\begin{equation} 
S_x^{j,k} = |j\rangle \langle k| + |k\rangle \langle j|. 
\end{equation} 

Its matrix elements are given by: 
\begin{equation} 
s_{mn} = 
\begin{cases} 1, & \text{if } (m,n) = (j,k) \text{ or } (k,j), \\ 0, 
& \text{otherwise}. 
\end{cases}
\end{equation} 

In a dense representation, one would store $d^{2n}$ elements, but only the two non-zero entries (and their indices) need to be stored with a sparse representation. Thus, while the state vector contains $d^n$ elements, each gate, especially for larger $d$ or operations involving many qudits, would require only storing a few elements instead of the full $d^{2n}$ entries.

Moreover, in simulations where gradient-based optimization is employed, the entire circuit (i.e., all layers of gates) must be held in memory to perform backpropagation. Even though dynamic circuit construction can reduce memory usage by only requiring the last layer to be stored for forward propagation, gradient computations necessitate access to all gate representations. Therefore, reducing the memory cost per gate via sparse representations becomes critical.

We investigated similar sparse representations for other quantum gates by identifying recurring patterns and symmetries. By directly constructing these sparse matrices, we avoid the overhead of first creating dense representations, which further reduces memory consumption during both forward and backward passes of the simulation. A more detailed derivation of the sparse representations can be found in Appendix \ref{appendix:2}.

\subsection{Code style}
\label{sec:codestyle}

QuForge supports multiple writing styles for building quantum circuits, accommodating different programming preferences and requirements. This flexibility aim to favor an algorithm-centric style, reminiscent of libraries like Qiskit \cite{qiskit2024} and Keras \cite{keras}, where circuits are constructed by sequentially stacking gates. Alternatively, it supports a machine learning-centric style, inspired by PyTorch, where circuits are defined through initialization and forward methods. Additionally, QuForge offers a hybrid style, which is particularly advantageous for data-centric problems. Each style presents unique advantages and disadvantages, making them suitable for specific issues and allowing users to choose the approach that best fits their needs.

The algorithm below demonstrates an example of constructing and simulating a quantum circuit with qudits using QuForge. First, we instantiate a circuit by defining the qudit dimension (\texttt{dim}), the total number of qudits (\texttt{wires}), and optionally, the device (\texttt{device}) on which to run the circuit. Next, we apply various quantum gates to the circuit, each with its properties specified through arguments. After setting up the gates, we prepare an initial state. QuForge allows state preparation by calling the \texttt{State} function and specifying the desired state. For instance, to prepare the qutrit state $|012\rangle$, we can call \texttt{State('0-1-2', dim-3)}.

To process the initial state, we pass it through the circuit by calling the circuit with the initial state as the input. The output corresponds to the final state after the circuit operations. Finally, to obtain the measurement history or probabilities, we call the \texttt{measure} function. This function returns a histogram and the probabilities of the measurement outcomes, providing the statistics of measuring the output state.

\begin{lstlisting}
import quforge.quforge as qf
import numpy as np

#Create a circuit
circuit = qf.Circuit(dim=[2,3,4], 
                     wires=3, 
                     device='cuda')

#Hadamard gate on all qudits
circuit.H(index=[0,1,2]) 

#RX gate on the first qudit with angle=pi/2
circuit.RX(angle=np.pi/2, index=[0])

#CNOT gate between the first and third qudit
circuit.CNOT(index=[0,2])

#Prepare state |013>
initial_state = qf.State('0-1-3', 
                         dim=[2,3,4], 
                         device='cuda')

#Run the circuit
output_state = circuit(initial_state)

#Measure the first two qudits, returns histogram, and probabilities
hist, prob = qf.measure(output_state, 
                        index=[0,1], 
                        dim=[2,3,4], 
                        wires=3, 
                        shots=1000)

\end{lstlisting}

In the current state of quantum computing devices, known as Noisy Intermediate-Scale Quantum (NISQ) technology \cite{Preskill_2018}, the number of qubits (or qudits) is insufficient for performing large-scale computations. This limitation is primarily due to noise, which induces cumulative errors that degrade the accuracy and reliability of computational results. Given these constraints, NISQ devices cannot address large-scale problems independently.

However, hybrid classical-quantum algorithms \cite{hybrid1, hybrid2} offer a promising approach to leverage the strengths of both classical and quantum computing for large-scale problems, such as image and text processing. These hybrid algorithms combine classical methods with quantum algorithms to optimize computational efficiency and performance. Typically, the classical component involves techniques to reduce the dimensionality of the problem, such as Principal Component Analysis (PCA) or neural networks. Compressing the data into a lower-dimensional representation makes the situation more manageable for quantum processing.

The quantum component then processes this compressed representation to achieve the desired objective. By operating on a smaller, more tractable data set, quantum algorithms can potentially outperform classical algorithms in specific tasks, despite the current limitations of NISQ devices. This hybrid approach mitigates the impact of noise and maximizes the use of available quantum resources.

The algorithm below illustrates an example of a hybrid classical-quantum algorithm structured for ease of use in machine learning applications. In this example, we define a class combining classical and quantum components, encapsulating them within a single object. This class includes an optimizer that manages the parameters of the algorithm, enabling their optimization.

First, we instantiate a hybrid model by defining its qudit dimension (\texttt{dim}) and the total number of qudits (\texttt{wires}). The class constructor initializes a quantum circuit, a rotation gate (\texttt{RZ}) that operates on all qudits, and a classical linear encoder. The initial state is also defined within the constructor. The \texttt{forward} method processes input data by encoding it using the classical linear layer, which produces parameters for the rotation gate. The state is then processed through the quantum circuit to produce the final output.

Next, we instantiate the hybrid model and an optimizer to optimize the model's parameters. During the training phase, we pass input data through the model to obtain the output state. We then compute the loss by comparing the output and target states. The optimizer updates the model parameters by computing the gradients and applying a gradient-based optimization step, effectively training the algorithm end-to-end.

\begin{lstlisting}
import quforge.quforge as qf
import quforge.cml as cml

class Hybrid(qf.Module):
    def __init__(self, dim, wires):
        super(Hybrid, self).__init__()

        self.circuit = qf.Circuit(dim=dim, wires=wires)
        self.circuit.RX()
        self.circuit.RY()
        self.circuit.RZ()
        self.circuit.CNOT()
        self.circuit.RX()
        self.circuit.RY()
        self.circuit.RZ()

        self.RZ = qf.RZ(
        dim=dim, 
        wires=wires,      
        index=range(wires)
        )

        self.encoder = cml.Linear(200, wires)

        self.initial_state = qf.State('0-0', dim=dim)

    def forward(self, x):
        parameters = self.encoder(x).flatten()
        state = self.RZ(self.initial_state, parameters)
        output = self.circuit(state)

        return output

#Instantiate the circuit and optimizer
model = Hybrid(dim=5, wires=2)
optimizer = qf.optim.Adam(model.parameters())

#Create a random normal vector as input
data = cml.randn((1, 200))

#Define a target state
target_state = qf.State('1-1', dim=5)

#Run and optimize the circuit
for epoch in range(100):
    output_state = model(data)

    loss = abs(target_state - output_state).sum()
    optimizer.zero_grad()
    loss.backward()
    optimizer.step()

\end{lstlisting}

\section{Results}
\label{sec:examples}

We demonstrate the feasibility of the QuForge library by implementing three distinct algorithms: Deutsch-Jozsa's algorithm, Grover’s algorithm, and a variational quantum circuit applied to two different problems.

\subsection{Deutsch-Jozsa algorithm}

The Deutsch-Jozsa algorithm \cite{Deutsch1992RapidSO} was one of the first quantum algorithms to demonstrate a quantum advantage. It determines whether a function $f(x)$ is constant, $f(x)=y$ for all $x$, or balanced, $f(x)=y_1$ for half of the inputs and $f(x)=y_2$ for the other half. Classically, evaluating deterministically such a function requires up to $2^{n-1}-1$ oracle queries. However, the Deutsch-Josza algorithm requires only a single query, thus offering an exponential reduction in complexity from $O(2^n)$ to $O(1)$.

In its quantum circuit representation, the algorithm begins with an initialized state where the register qubits are prepared in the $|0\rangle$ state and the evaluation qubit is prepared in the state $|1\rangle$. Hadamard gates are then applied to all qubits. Subsequently, a quantum oracle evaluates the inputs, representing the function under test. Another set of Hadamard gates is applied to the register qubits before measurement. The function is deemed constant if all qubit measurements equal zero; otherwise, it is balanced.

\begin{figure}[H]
    \centering
    \centering
    \includegraphics[width=1\linewidth]{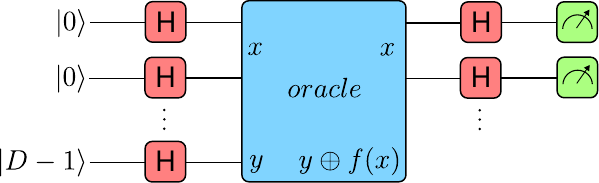}
    \caption{Quantum circuit diagram representing the generalized Deutsch-Jozsa algorithm for qudits. This configuration uses the state $|D-1\rangle$ for the last qudit and employs generalized Hadamard gates. The circuit tests whether a given function, represented as a black box in the diagram, is constant or balanced across $d$-dimensional states (dits).}
    \label{fig:deutsch}
\end{figure}

The extension of this algorithm to qudits is straightforward \cite{deutsch_qudit1, deutsch_qudit2}, as illustrated in Figure \ref{fig:deutsch}. Instead of using the state $|1\rangle$ for the last qubit, we use the qudit state $|D-1\rangle$, where $D$ represents the dimension of the qudit. By measuring the register qudits, information is obtained in dits rather than bits, increasing the classical complexity to $D^{n-1}-1$ evaluations, while the quantum complexity remains constant.

We provide sample code using the library to illustrate the practical implementation of the Deutsch-Josza algorithm for qudits. The code below demonstrates how to set up the circuit for the algorithm, define an oracle function that can be either constant or balanced, and execute the necessary quantum operations to determine the function's nature. By leveraging the flexibility of qudits, the circuit employs generalized Hadamard gates and measures the system in dits, reflecting the dimensionality $D$ of the qudits involved.

\begin{lstlisting}
import quforge.quforge as qf

dim = 3 #dimension of the qudit
wires = 4 #number of qudits

# Define oracle
def oracle(model, mode='constant'):
    if mode == 'constant':
        model.X(index=[wires-1])
    else:
        model.CNOT(index=[2,3])
    return model

# Define circuit
circuit = qf.Circuit(dim=dim, 
                     wires=wires, 
                     device=device)
circuit.H(index=range(wires))

# Apply oracle
oracle(circuit, mode='constant')

# Hadamard on the first N-1 qudits
circuit.H(index=range(wires-1))

# Initial state
state = ''
for i in range(wires-1):
    state += '0-'
state += '%i' % (dim-1)
state = qf.State(state, dim=dim, device=device)

# Apply the initial state in the circuit
output = circuit(state)

# Measure the first N-1 qudits
histogram, p = qf.measure(output, 
                          index=range(wires-1), 
                          dim=dim, 
                          wires=wires, 
                          shots=1024)

if p[0] == 1:
    print('The function is constant')
else:
    print('The function is balanced')
\end{lstlisting}

\subsection{Grover's algorithm}

Grover's algorithm \cite{grover} is important in the field of quantum computing, providing an optimized search mechanism for unstructured data. Classical search algorithms operate with a complexity of $O(N)$, where $N$ represents the total number of items. Grover's algorithm achieves a quadratic speed-up, demonstrating a complexity of $O(\sqrt{N})$. This improvement means the algorithm can locate items within a database faster, on average.

Fundamentally, Grover's algorithm is based on an oracle mechanism. It operates in two main phases: the oracle and diffusion stages. In the oracle phase, the desired item is marked by flipping its phase. Subsequently, the diffusion process amplifies the amplitude of this marked state, thereby increasing the probability that the correct item will be selected upon measurement. This sequence is iteratively repeated until the amplitude of the marked state is maximized. It is crucial to note that excessively repeating these iterations can inadvertently decrease the amplitude of the marked state due to the averaging effect introduced by the diffusion operator. Therefore, determining the optimal number of iterations is essential to maximize the effectiveness of the search.

\begin{figure}[H]
    \centering
    \centering
    \includegraphics[width=1\linewidth]{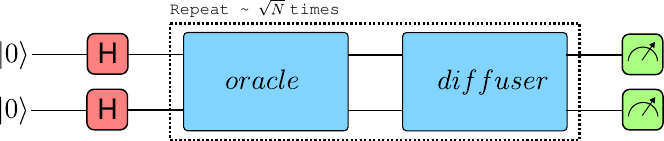}
    \caption{Quantum circuit diagram representing the generalized Grover's algorithm. The circuit has an iterative part repeated at the order of $\sqrt{N}$ times. The oracle is responsible for flipping the phase of the marked state, while the diffusion step amplifies its amplitude.}
    \label{fig:grover}
\end{figure}

The core principles largely remain unchanged when extending Grover's algorithm to qudits \cite{grover_qudit1, grover_qudit2}, as shown in Figure \ref{fig:grover}. However, using qudits allows for a more extensive database capacity, albeit at the cost of an increased number of necessary iterations. A critical step involves achieving an equal superposition of all states before applying the algorithm itself, which can be accomplished by using a generalized Hadamard gate on the $|0\rangle$ state.

To demonstrate the implementation of Grover's algorithm for qudits using the QuForge library, the following code outlines the construction of a quantum circuit that searches for a marked state within an unstructured database. The code initializes the circuit by creating an equal superposition of all possible states using generalized Hadamard gates. It then iteratively applies the oracle, which marks the target state by flipping its phase, followed by the Grover diffusion step, which amplifies the probability amplitude of the marked state. By carefully choosing the number of iterations, the algorithm efficiently enhances the likelihood of identifying the marked state upon measurement.

\begin{lstlisting}
import quforge.quforge as qf

dim = 3 #dimension of the qudit
wires = 2 #number of qudits

# Define oracle
def oracle(circuit, marker): 
    state = qf.State(marker, dim=dim, device=device)
    target_state = state @ state.conj().T

    U = qf.eye(dim**wires, device=device) - 2 * target_state
    circuit.U(matrix=U, index=[0,1])

# Define the Grover diffusion step
def grover_diffusion(circuit, state): 
    U = 2 * state @ state.conj().T - qf.eye(dim**wires, device=device)
    circuit.U(matrix=U, index=[0,1])

# Apply Hadamard on each qudit
input_state = qf.State('0-0', 
                       dim=dim, 
                       device=device)
H = qf.H(dim=dim, 
         wires=wires, 
         index=[0,1], 
         device=device)

state = H(input_state)

# Create circuit
circuit = qf.Circuit(dim=dim, 
                     wires=wires, 
                     device=device)

# Apply the oracle. 
# In this example, we want to find the state |22>
oracle(circuit, '2-2') 

# Apply the diffusion
grover_diffusion(circuit, state)

# Apply the circuit on the initial state
output = circuit(state)

# Measure the first N-1 qudits
histogram, p = qf.measure(output, 
                          index=register, 
                          dim=dim, 
                          wires=wires,
                          shots=10000
                          )
                          
\end{lstlisting}

\subsection{Variational quantum algorithm}

Variational Quantum Algorithms (VQAs) \cite{Cerezo_2021} are among the principal quantum algorithms used in machine learning. These algorithms optimize parameterized circuits to minimize a loss function, achieving a specific objective. The circuits consist of parameterized gates, typically rotation gates with adjustable angles, that are iteratively modified to produce the desired outcomes. Generally, VQAs are optimized using gradient-based methods, where inference is performed by a quantum computer (or simulator), and the optimization is carried out classically.

Extending VQAs to qudits offers multiple advantages. First, they provide a more nuanced representation of data as they are not constrained by binary representations, allowing for encoding more information with the same number of qudits. Additionally, qudits enable more expressive output representations, effectively compressing more information into fewer qudits.

In this study, we demonstrate the application of qudit-based VQAs to two classification problems using the Iris and MNIST datasets, highlighting the advantages of qudits over qubits in specific scenarios.

\begin{figure}[t]
    \centering
    \centering
    \includegraphics[width=1\linewidth]{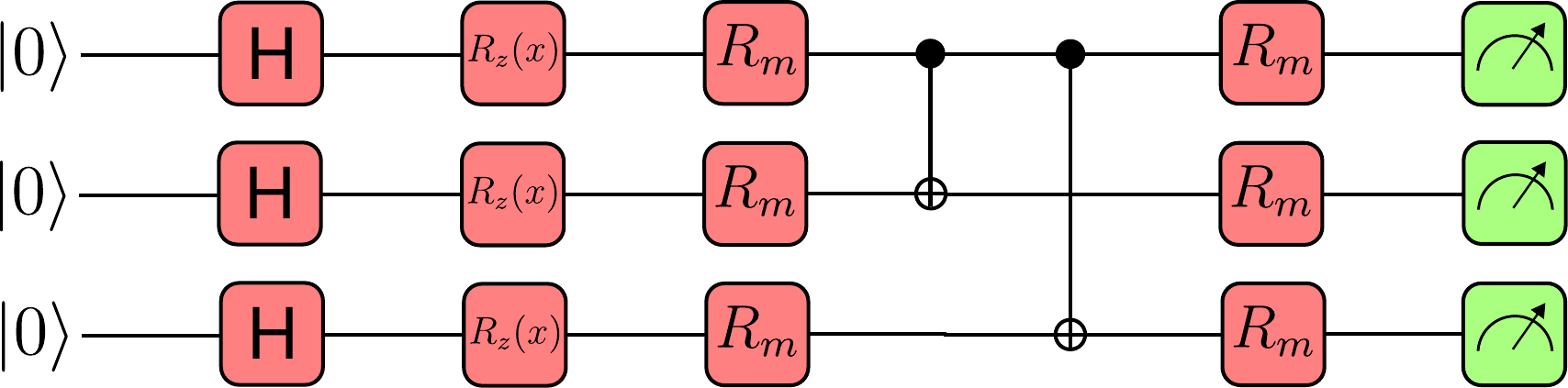}
    \caption{General quantum variational circuit chosen for this study. Hadamard gates are applied to the zero state before feature encoding of the data to prepare the input state for the algorithm. Then, several rotation gates are applied, followed by CNOT gates, and a second set of rotation gates are applied before measurement.}
    \label{fig:qva}
\end{figure}

Figure \ref{fig:qva} depicts the circuit configuration used in both classification tasks. The process initiates with all qudits in the zero state, represented as $|0\rangle^{\otimes N}$. Each qudit undergoes transformation via a generalized Hadamard gate, preparing the state for feature encoding. This encoding involves setting the angles of \texttt{RZ} rotation gates based on the input data features. Subsequently, the primary VQA circuit, which consists of a series of rotation gates, is employed. This is followed by applying generalized \texttt{CNOT} gates, which establish entanglement by coupling the first qudit with each subsequent qudit. The circuit concludes with a final series of rotation gates before measurement.

The selection of rotation gates, denoted as \texttt{R$_m$}, is contingent upon the dimensionality of the qudits and can be configured in various forms. We determine the index $m$ corresponding to each circuit configuration as outlined below:

\begin{itemize}
    \item  The first configuration, \texttt{R$_1$}, applies a sequence of rotation gates along the $z$, $y$, and $x$ axes to each qudit, restricted to the indexes $\{0,1\}$ for each gate, formally represented as:
\begin{equation}
R_1 |\psi \rangle = \prod_{i=x,y,z} R_i^{0,1} |\psi\rangle = R_z^{0,1}R_y^{0,1}R_x^{0,1} |\psi\rangle
\end{equation}

    \item In the second configuration, \texttt{R$_2$} rotation gates are applied across all axes with the index $j=0$ consistently maintained, while the index $k$ varies for the axes $x$ and $y$, and index $j$ varies for the axis $z$, as represented by:
\begin{equation}
R_2 |\psi \rangle = \prod_{i=x,y,z} \prod_{k} R_i^{0,k} |\psi\rangle
\end{equation}

    \item The third configuration, \texttt{R$_3$}, extends the application of rotation gates along all axes, and combines all possible rotations by varying the indices $j$ and $k$, as described by:
\begin{equation}
R_3 |\psi \rangle = \prod_{i=x,y,z} \prod_{j,k} R_i^{j,k} |\psi\rangle
\end{equation}

\end{itemize}

We provide a code example using the library to further explore the implementation of qudit-based Variational Quantum Algorithms. The code defines a quantum variational circuit designed explicitly for classification tasks, featuring an initial state preparation through generalized Hadamard gates, followed by feature encoding with \texttt{RZ} rotation gates. The primary variational circuit consists of multiple layers of rotation gates along different axes and generalized \texttt{CNOT} gates that establish entanglement among the qudits. This setup allows for flexible circuit configurations based on the chosen rotation gates, as described in the preceding sections. By adjusting the circuit parameters through a classical optimization loop, the model learns to minimize the loss function, thus achieving the desired classification outcomes. The following code outlines the circuit structure and demonstrates the process of initializing, encoding, and evolving the qudit states within the variational framework.

\begin{lstlisting}
import quforge.quforge as qf

class Circuit(qf.Module):
    def __init__(self, dim, wires, device):
        super(Circuit, self).__init__()

        self.dim = dim
        self.init = qf.H(dim=dim, 
                         index=range(wires), 
                         device=device)
        self.encoder = qf.RZ(dim=dim, 
                             index=range(wires), 
                             device=device)

        self.circuit = qf.Circuit(dim=dim, 
                                  wires=wires, 
                                  device=device)
        self.circuit.RX(j=0, k=1, index=range(wires))
        self.circuit.RX(j=1, k=2, index=range(wires))
        self.circuit.RX(j=0, k=2, index=range(wires))
        self.circuit.RY(j=0, k=1, index=range(wires))
        self.circuit.RY(j=1, k=2, index=range(wires))
        self.circuit.RY(j=0, k=2, index=range(wires))
        self.circuit.RZ(j=1, index=range(wires))
        self.circuit.RZ(j=2, index=range(wires))
        self.circuit.CNOT(index=[0,1])
        self.circuit.CNOT(index=[0,2])
        self.circuit.CNOT(index=[0,3])
        self.circuit.RX(j=0, k=1, index=range(wires))
        self.circuit.RX(j=1, k=2, index=range(wires))
        self.circuit.RX(j=0, k=2, index=range(wires))
        self.circuit.RY(j=0, k=1, index=range(wires))
        self.circuit.RY(j=1, k=2, index=range(wires))
        self.circuit.RY(j=0, k=2, index=range(wires))
        self.circuit.RZ(j=1, index=range(wires))
        self.circuit.RZ(j=2, index=range(wires))

        self.initial_state = qf.State('0-0-0-0', 
                                      dim=dim, 
                                      device=device)

    def forward(self, x):
        y = self.init(self.initial_state)
        y = self.encoder(y, param=x)
        y = self.circuit(y)

        return y

circuit = QVA(dim=3, wires=4)
\end{lstlisting}

\subsubsection{Iris dataset}

The Iris Flower dataset \cite{misc_iris_53} comprises observations from 50 individual flowers distributed across three distinct species: Iris setosa, Iris virginica, and Iris versicolor. Each flower is described by four phenotypic attributes: Petal Length, Petal Width, Sepal Length, and Sepal Width.

In our approach, we utilize a quantum circuit with four qudits, where each qudit is responsible for phase encoding a specific attribute. Given the classification of the dataset into three species, we implement a qubit and a qutrit for categorical encoding. The qubit employs one-hot encoding to represent the three classes, defined by the target states as $|100\rangle$, $|010\rangle$, and $|001\rangle$, respectively, measured on the first three qubits. Concurrently, the qutrit is also used for one-hot encoding of the species classes, focusing on exploring the properties of high-dimensionality intrinsic to qudits. Here, we measure only the first qutrit, designating the target states as $|0\rangle$, $|1\rangle$, or $|2\rangle$.

The circuit is configured as described before, with the rotation gates chosen from the \texttt{R$_3$} design. The dataset is partitioned into subsets, with 40 data points designated for training and the remaining $10$ are used for testing purposes. We conducted training across five different initializations to evaluate the mean classification accuracy.

\begin{figure}
    \centering
    \begin{subfigure}{0.49\textwidth}
    \centering
        \includegraphics[width=1\linewidth]{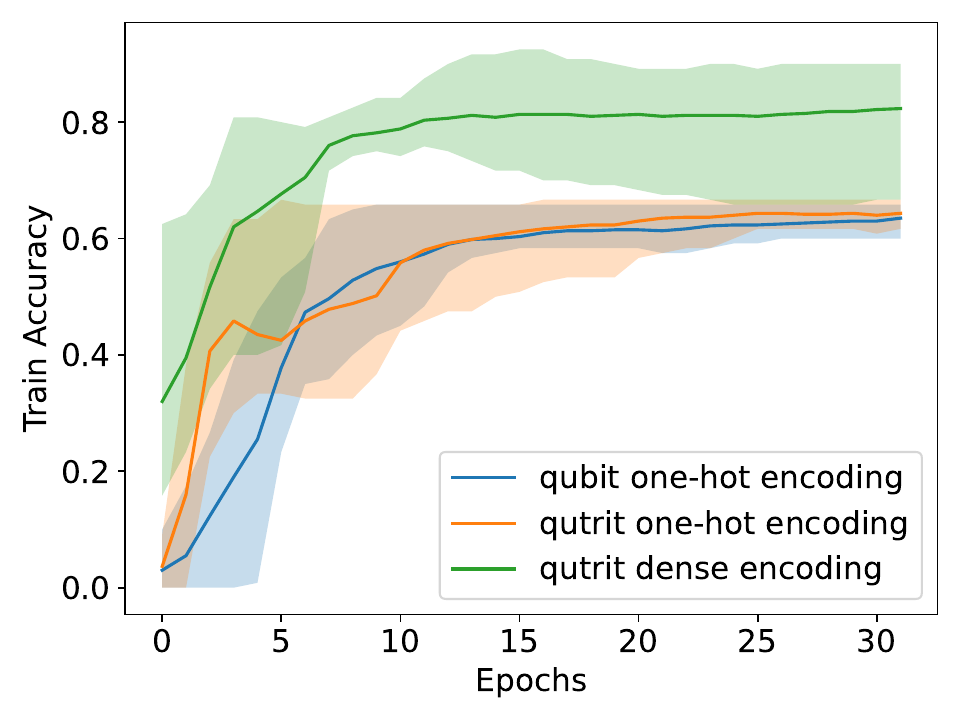}
        \caption{}
        \label{}
    \end{subfigure}%
    
    \begin{subfigure}{0.49\textwidth}
    \centering
        \includegraphics[width=1\linewidth]{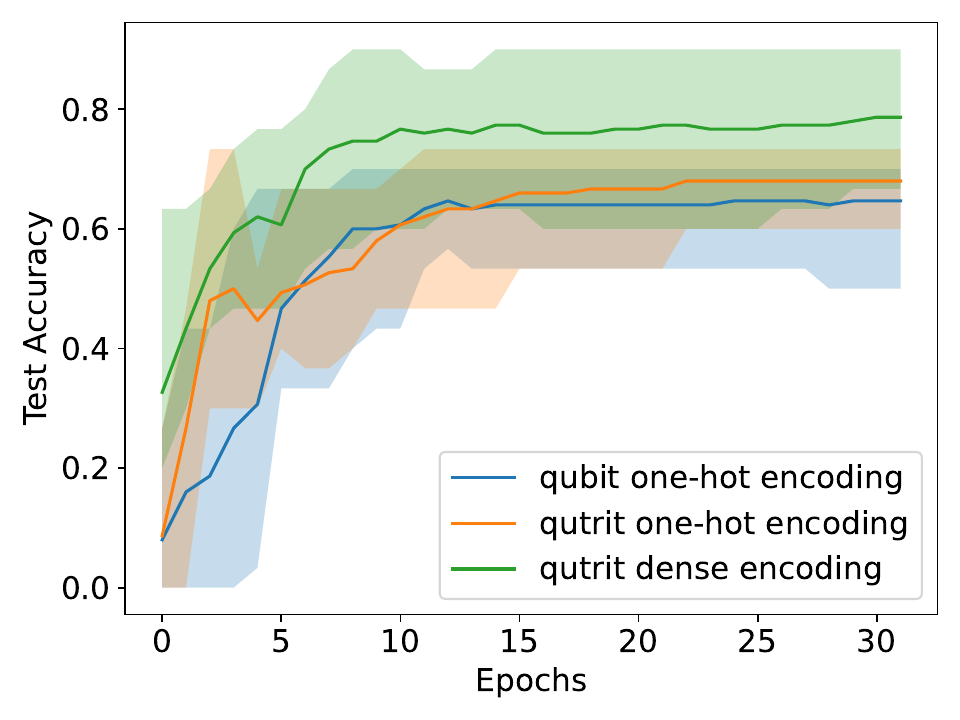}
        \caption{}
        \label{}
    \end{subfigure}
    \caption{\justifying Mean accuracy of the Iris dataset for (a) training data and (b) test data for different dimensions of qudits and types of encoding. The shaded regions indicate the lowest and highest accuracy obtained for each run.}
    \label{fig:iris}
\end{figure}

Figure \ref{fig:iris} illustrates the comparative results from each model configuration. The results indicate that employing a dense encoding scheme for the qutrits yields superior accuracy compared to the alternate models. This enhancement suggests the potential benefits of leveraging the higher dimensionality of qudits in quantum variational algorithms.

\subsubsection{MNIST dataset}

The MNIST dataset \cite{lecun2010mnist} comprises labeled images of handwritten digits ranging from 0 to 9, each represented in grayscale with a resolution of 28x28 pixels. Given the considerable dimensionality presented by the 784-pixel input space for quantum simulation, our methodology adopts a hybrid approach aimed at dimensionality reduction to minimize the requisite number of qudits for computational processing. This process involves using a classical linear layer model, comprising a single layer of neurons that condenses the 784-dimensional vector representation of each image into a compact form represented by the number of qudits available. Subsequently, phase encoding is applied to the output of each neuron, thereby preparing a quantum state for training a variational circuit.

Considering the categorization of the dataset into ten distinct classes, our model employs a ten-dimensional qudit to represent each class. This classification is quantified by measuring the state of the first qudit. For comparative analysis with traditional qubit-based circuits, we implement a circuit configured with ten qubits, utilizing a one-hot encoding scheme to represent each class.

In this study, a subset comprising 1000 randomly selected images serves as the training set, while an additional set of 500 randomly chosen images is designated for testing purposes. The efficacy of the proposed model is evaluated by calculating the mean accuracy across both datasets.

\begin{figure}
    \centering
    \begin{subfigure}{0.49\textwidth}
    \centering
        \includegraphics[width=1\linewidth]{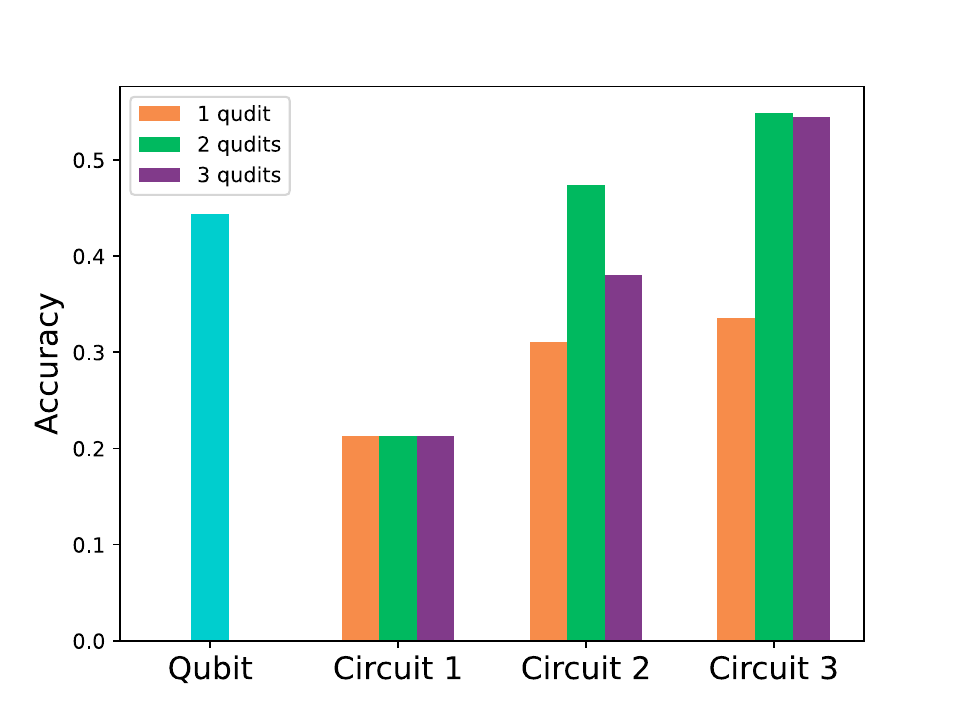}
        \caption{}
        \label{}
    \end{subfigure}%
    
    \begin{subfigure}{0.49\textwidth}
    \centering
        \includegraphics[width=1\linewidth]{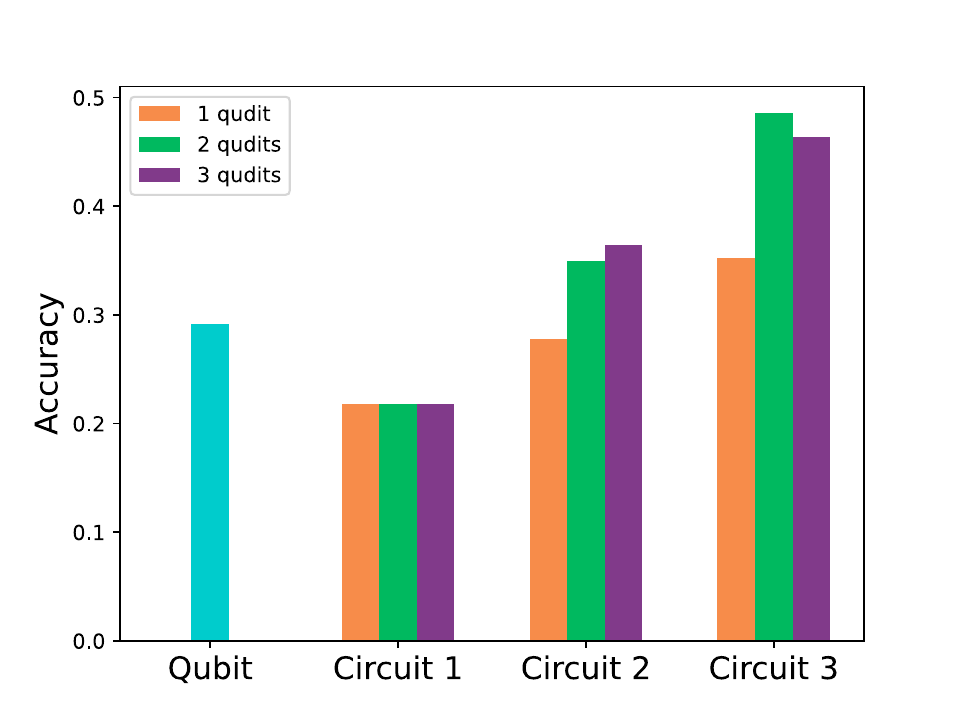}
        \caption{}
        \label{}
    \end{subfigure}
    \caption{\justifying Mean accuracy for (a) training data and (b) test data across various circuits and qudit numbers. For shallow circuits (Circuit 1), qudits exhibit lower accuracy than qubits. However, with increased rotation gates, qudit models (Circuits 2 and 3) surpass qubits in accuracy.}
    \label{fig:mnist}
\end{figure}

We applied the three designs of rotation gates to test different possible combinations. As depicted in Figure \ref{fig:mnist}, our analysis reveals that utilizing at least two qudits not only achieves but exceeds the accuracies obtainable with a ten-qubit circuit. This observation suggests that qudits can offer superior representational efficiency with fewer quantum bits, highlighting their potential in quantum computations with qudits.

\subsection{Performance metrics}

Our performance evaluation measures two distinct time metrics: initialization and execution times. The initialization time refers to the time required to construct the quantum circuit after its specification. This includes the instantiation of all gate objects, the pre-computation of their matrix representations, and any device-specific setup. For non-parametric gates (such as \texttt{Hadamard}, \texttt{CNOT}, and \texttt{X} gates), the initialization time comprises the construction of the dense (or sparse) matrix representations. Parametric gates, such as the \texttt{RX} gate, include the setup and allocation of the rotation parameter and the construction of any static gate components. This time metric is critical when performing gradient-based optimization during classical simulation, as the entire circuit, including all gate matrices, must be stored in memory for backpropagation.

To ensure robustness in the performance analysis, we conduct 20 independent trials for each configuration and compute the mean and standard deviation of the measured times. This approach enables us to quantify the sensitivity of the library’s performance to variations in execution conditions, analysing potential fluctuations arising from hardware load or architectural differences.

We focus our evaluation on quantum variational circuits, as depicted in Figure \ref{fig:qva}, as they are very general for quantum machine learning. However, it is important to note that the specific choice of circuit architecture and the types of quantum gates employed may influence initialization and execution. Therefore, while quantum variational circuits provide a solid benchmark for performance assessment, other circuit types may exhibit different timing characteristics depending on their structural complexity and gate set.

\begin{figure*}[ht]
    \centering
    \begin{subfigure}{0.49\textwidth}
    \centering
        \includegraphics[width=1\linewidth]{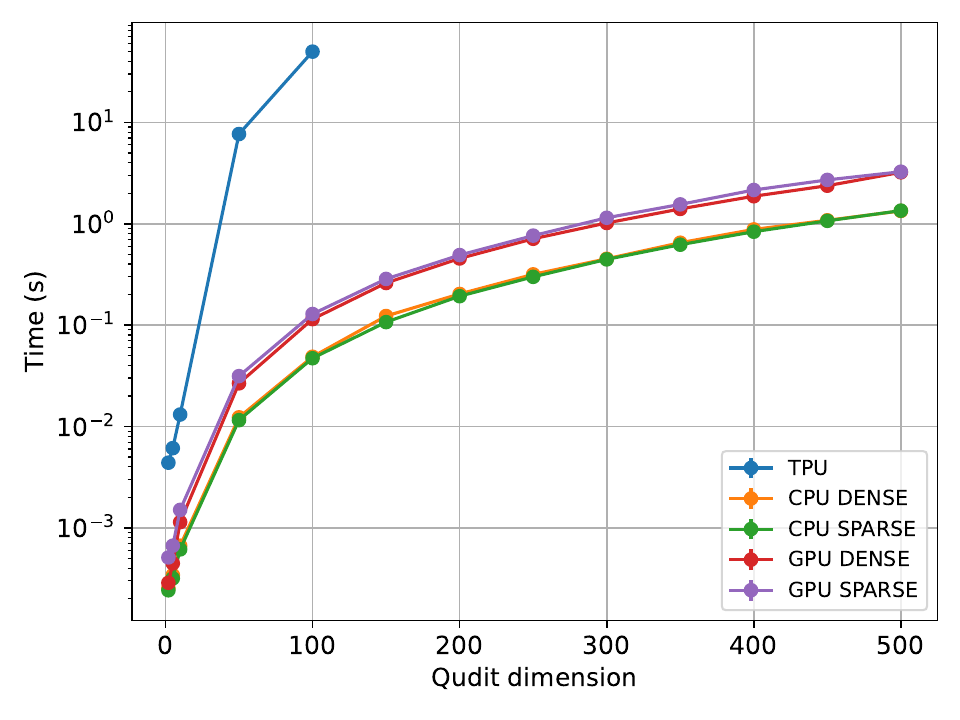}
        \caption{}
        \label{}
    \end{subfigure}%
    \begin{subfigure}{0.49\textwidth}
    \centering
        \includegraphics[width=1\linewidth]{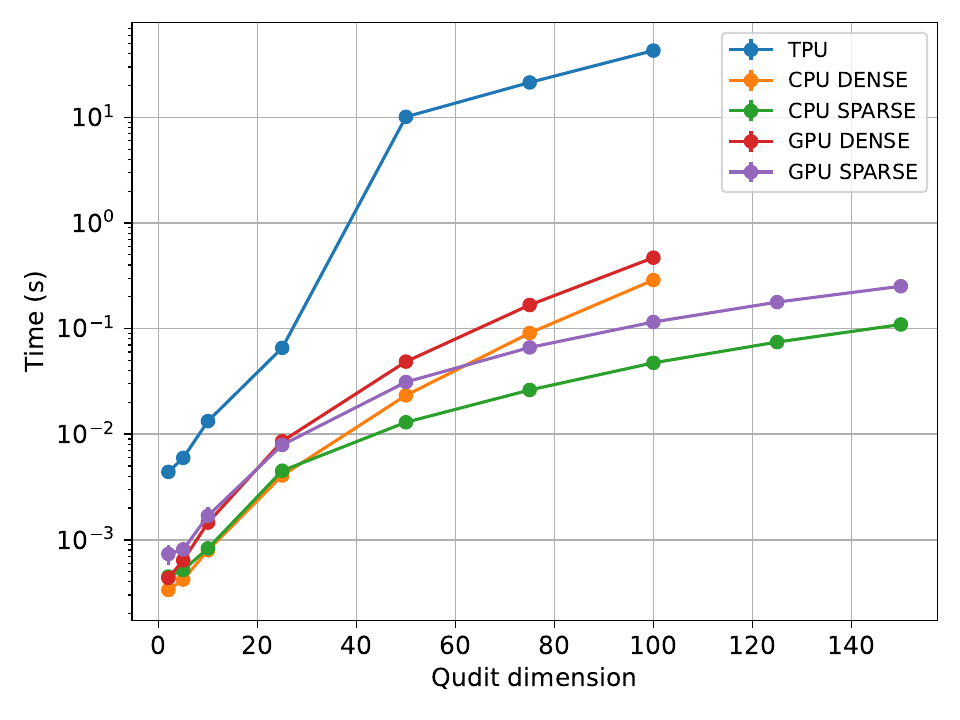}
        \caption{}
        \label{}
    \end{subfigure}%

    \begin{subfigure}{0.49\textwidth}
    \centering
        \includegraphics[width=1\linewidth]{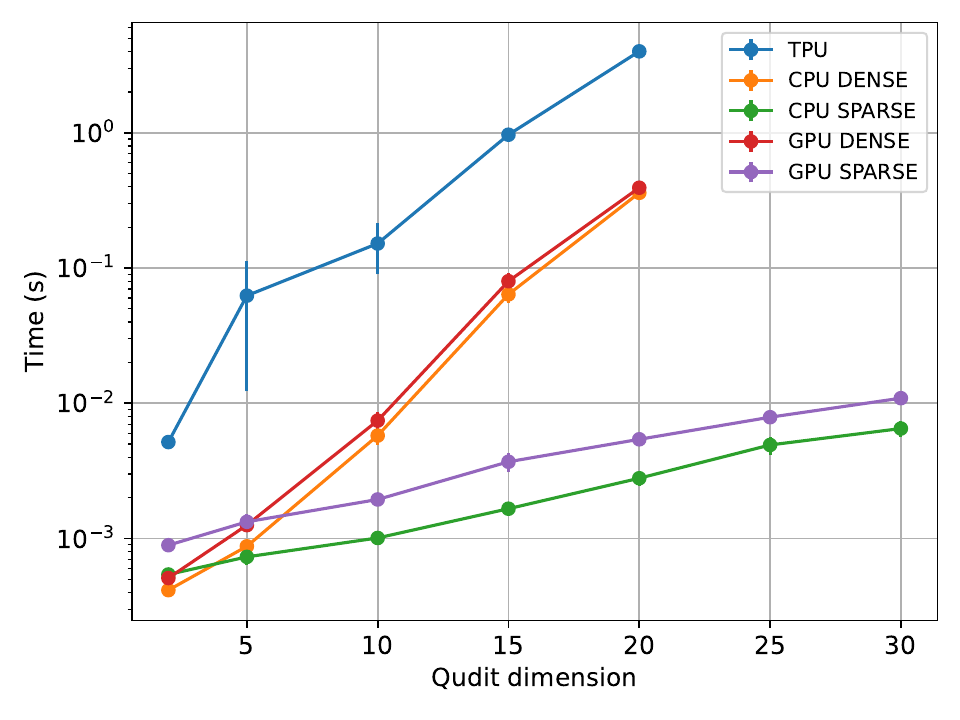}
        \caption{}
        \label{}
    \end{subfigure}%
    \begin{subfigure}{0.49\textwidth}
    \centering
        \includegraphics[width=1\linewidth]{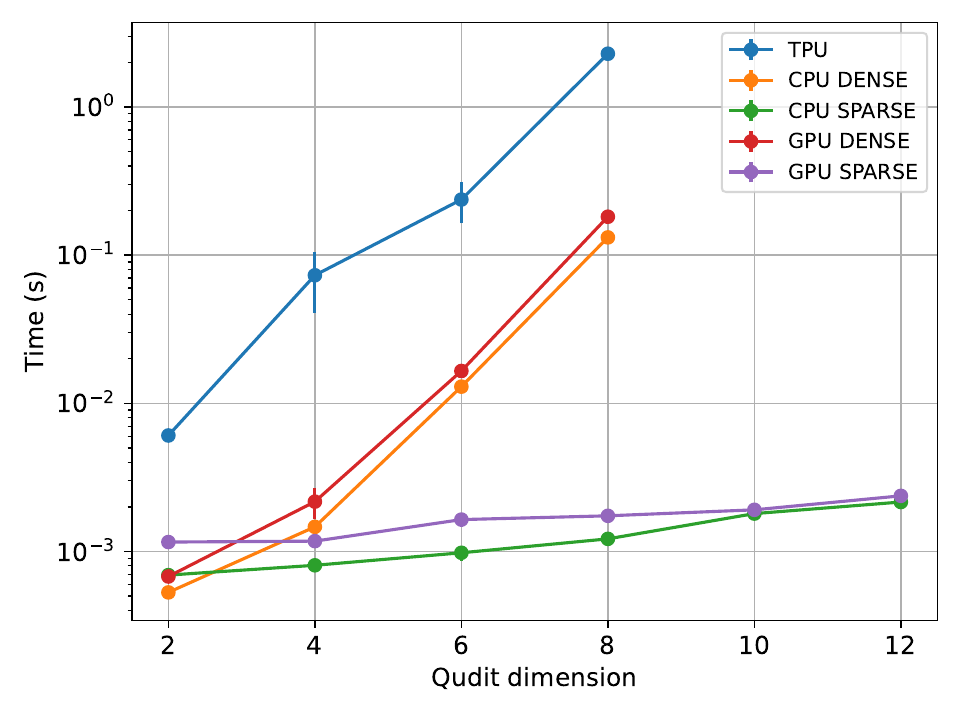}
        \caption{}
        \label{}
    \end{subfigure}%

    \caption{\justifying Initialization time as a function of the number of qudits and qudit dimensions: (a) 1 qudit, (b) 2 qudits, (c) 3 qudits, and (d) 4 qudits. CPU and GPU performance is compared for sparse and dense matrix representations, and TPU measurements are shown for reference. Sparse matrices consistently reduce initialization time as qudit count grows, particularly on the CPU, while TPU initialization overhead remains significantly higher.}
    \label{fig:measuretime_init}
\end{figure*}

Figure \ref{fig:measuretime_init} shows the initialization time of the qudit simulation library for CPU, GPU, and TPU devices across different qudit counts and dimensions.  The observed behavior demonstrates that initialization time is sensitive to both the number of qudits and the dimensionality of the qudit space. The performance difference between the CPU and GPU configurations, with sparse and dense gate representations, is minimal for a single qudit. Sparse matrix representations show a slight advantage in initialization time over dense matrices, although this difference remains marginal. By contrast, the TPU initialization curve lies an order of magnitude above both CPU and GPU for all but the smallest dimensions, reflecting the overhead of device setup and compilation on TPU hardware.

However, as the number of qudits increases, the performance divergence becomes more pronounced. When simulating multiple qudits, sparse matrix representations consistently outperform dense representations, particularly on the CPU. Dense matrices on the CPU maintain an edge in performance only at very low qudit dimensions. For configurations with three and four qudits, the sparse matrix representation significantly reduces initialization time compared to its dense counterpart, demonstrating clear scalability advantages in larger systems. Although TPUs enable high-throughput matrix operations, their initialization overhead remains the dominant cost across all multi-qudit benchmarks.

Notably, results suggest that sparse matrix operations on the GPU could potentially offer faster initialization times for systems with four qudits at high dimensions. However, this advantage was not fully realized in our experiments due to memory limitations inherent to high-dimensional simulations on GPU hardware. It is important to highlight that sparse representations allow simulations at higher qudit dimensions, thanks to their reduced memory relative to dense representations. This reduced memory consumption is particularly beneficial when scaling up the dimensionality of qudits, as it mitigates the memory overhead typically associated with dense quantum gate representations.

\begin{figure*}
    \centering
    \begin{subfigure}{0.49\textwidth}
    \centering
        \includegraphics[width=1\linewidth]{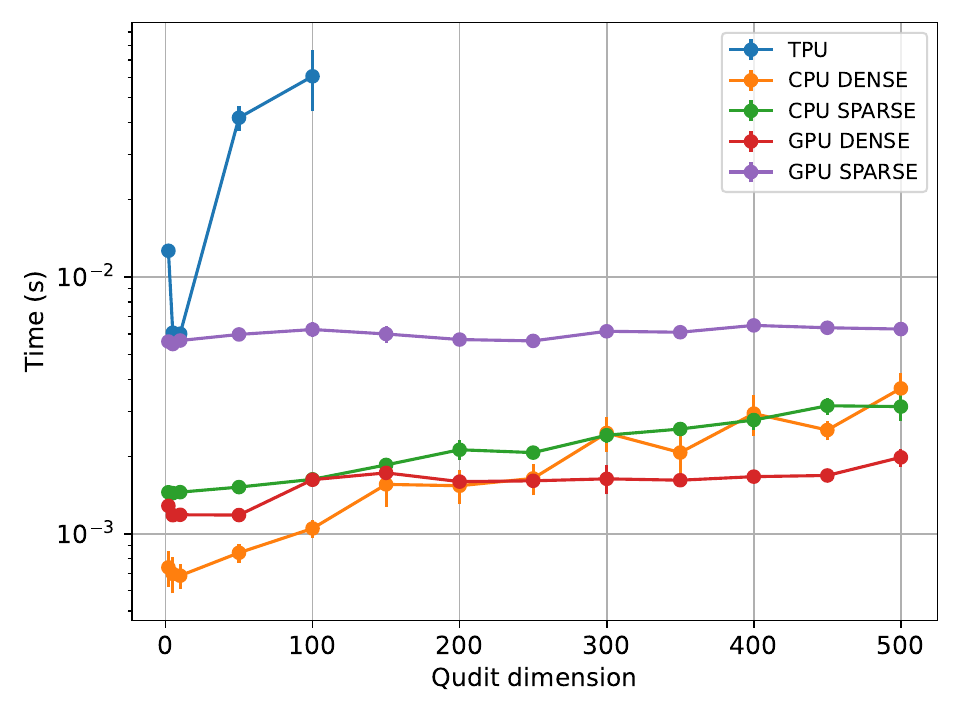}
        \caption{}
        \label{}
    \end{subfigure}%
    \begin{subfigure}{0.49\textwidth}
    \centering
        \includegraphics[width=1\linewidth]{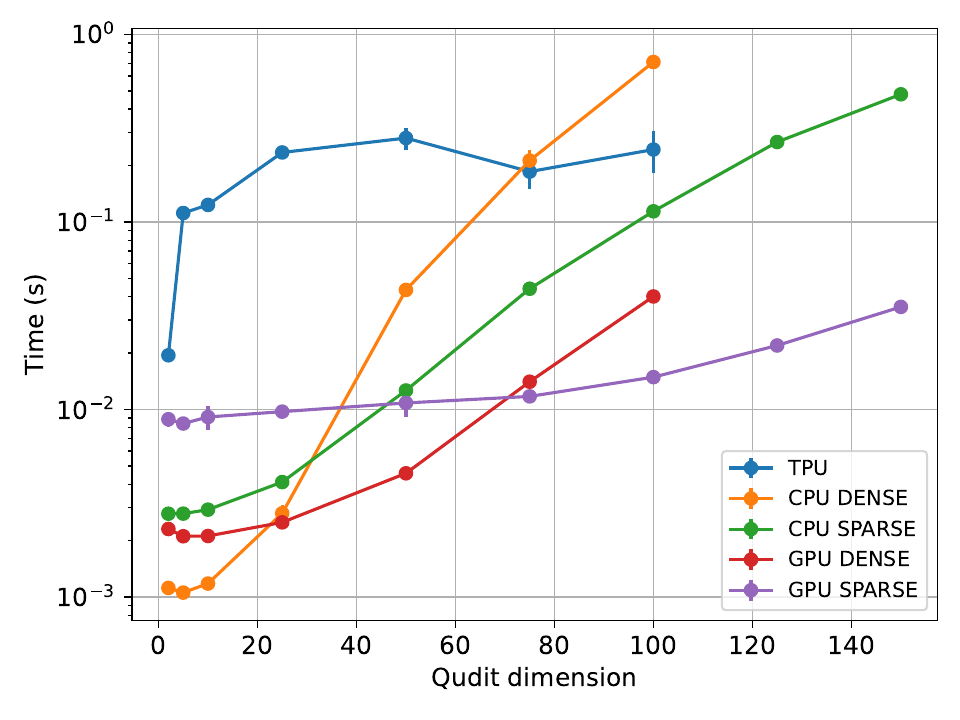}
        \caption{}
        \label{}
    \end{subfigure}%

    \begin{subfigure}{0.49\textwidth}
    \centering
        \includegraphics[width=1\linewidth]{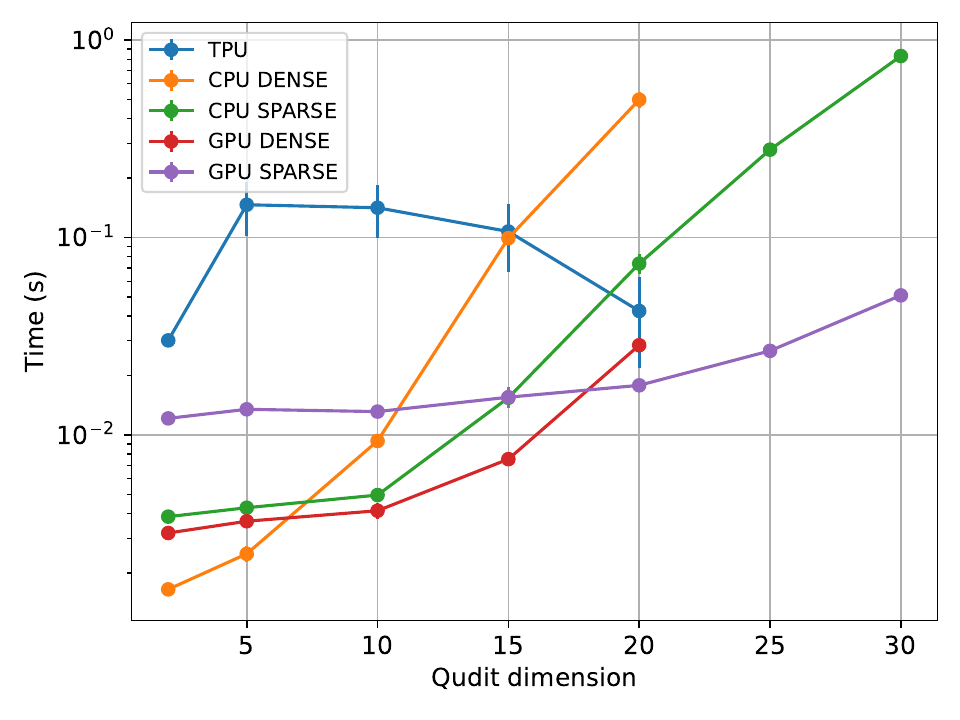}
        \caption{}
        \label{}
    \end{subfigure}%
    \begin{subfigure}{0.49\textwidth}
    \centering
        \includegraphics[width=1\linewidth]{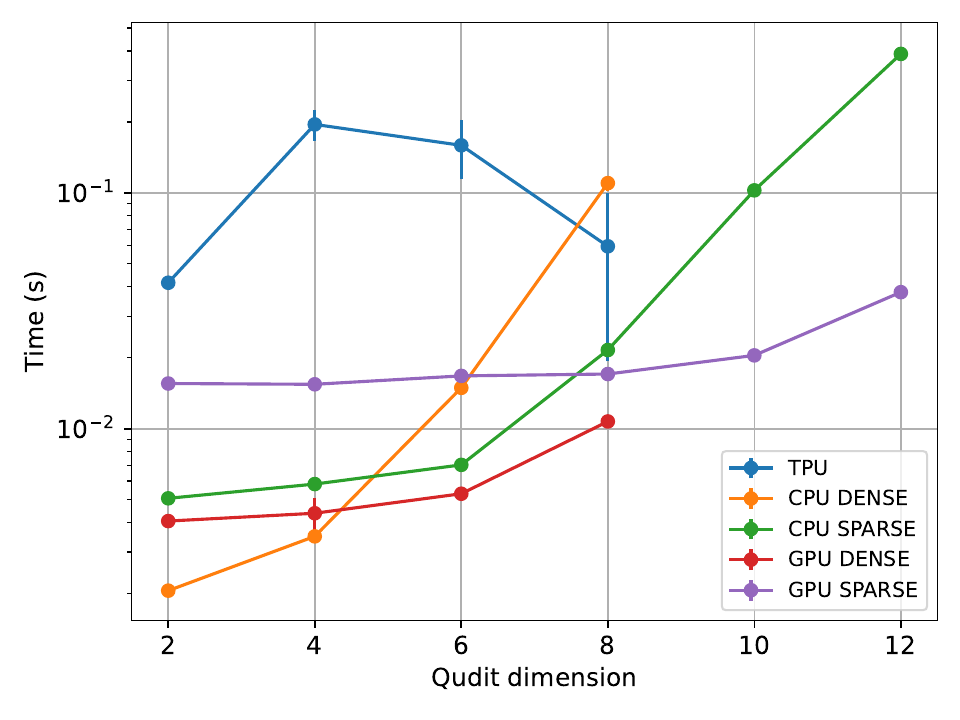}
        \caption{}
        \label{}
    \end{subfigure}%

    \caption{\justifying Execution time as a function of the number of qudits and qudit dimensions: (a) 1 qudit, (b) 2 qudits, (c) 3 qudits, and (d) 4 qudits, comparing CPU and GPU performance with sparse and dense matrix representations, and including TPU measurements. CPU dense matrices are fastest at low dimensions, but as dimensionality and qudit count increase, GPU sparse representations show a performance advantage, while TPU execution performance lags due to data transfer and compilation overhead.}
    \label{fig:measuretime_run}
\end{figure*}

Figure \ref{fig:measuretime_run} presents the execution time of quantum circuits for CPU, GPU, and TPU devices as qudit count and dimensions vary. The CPU with dense matrix representations emerges as the fastest configuration for a single qudit at low dimensions. However, the sparse CPU representation becomes more advantageous at high dimensions, whereas the sparse GPU configuration is the slowest. This slowdown in the GPU sparse case for a single qudit is likely due to the relatively low sparsity of the matrices, which diminishes the benefits of sparse matrix operations in such configurations. The TPU execution curve lies between the GPU and CPU dense lines for small qudit counts, but as dimensionality increases, TPU performance degrades relative to GPU sparse, reflecting a trade-off between high degrees of parallelism and data transfer overhead on TPU.

As the number of qudits increases, a clear trend emerges: sparse matrix representations on the GPU become the fastest at higher dimensions. The CPU dense configuration remains the most rapid for low dimensions, particularly at the lower end of the dimensional scale. However, as the dimensionality increases, both the GPU and sparse representations outperform dense CPU configurations, highlighting the efficiency gains in handling larger, more complex qudit systems with sparse matrices and GPU devices. As observed in the initialization time analysis, sparse matrix representations also enable the simulation of qudits at higher dimensions due to their significantly reduced memory requirements compared to dense matrices. This efficiency allows for the simulation of more complex systems, providing an advantage for high-dimensional quantum simulations.

\section{Conclusion}
\label{sec:conc}

This article has introduced QuForge, a Python library designed explicitly to simulate qudit-based quantum circuits. QuForge addresses critical gaps in quantum computing by offering a versatile platform that provides customizable quantum gates for arbitrary qudit dimensions. Leveraging the computational capabilities of modern frameworks such as PyTorch, QuForge enables seamless execution on diverse hardware platforms, including GPUs and TPUs, significantly enhancing simulation speed and scalability.

A distinguishing feature of QuForge is its ability to construct quantum circuits as differentiable graphs, which is particularly advantageous for quantum machine learning and hybrid classical-quantum algorithms. By reducing coding overhead and abstracting the complexities of these systems, QuForge allows researchers to focus on algorithm development and experimentation. Additionally, the implementation of sparse matrix operations substantially improves memory efficiency and computational performance, facilitating the simulation of larger and more complex quantum systems.

Nevertheless, limitations remain. QuForge is currently constrained to classical simulations, reflecting the predominant focus on qubit-based implementations in existing quantum hardware. The library does not yet support the decomposition of qudit operations into equivalent qubit operations, a feature that represents an active area of research and future development. As the field of qudit hardware matures, these functionalities could be integrated into subsequent versions.

QuForge aspires to be a tool for advancing qudit-based quantum research. By lowering barriers to exploration and experimentation, this library aims to catalyze innovation in quantum technologies, supporting the scientific community in unlocking the full potential of high-dimensional quantum systems.

\section*{Data availability}

The code for the QuForge library, along with the examples in this article, can be found at \url{https://github.com/tiago939/QuForge}.

\section*{Acknowledgments}

This work was supported by the S\~ao Paulo Research Foundation (FAPESP),
Grant No. 2023/15739-3, the Coordination for the Improvement of Higher Education Personnel (CAPES), Grant No. 88887.829212/2023-00, the National Council for Scientific and Technological Development (CNPq), Grants No. 309862/2021-3, No.
409673/2022-6, and No. 421792/2022-1, and by the National Institute for the Science and Technology of Quantum Information (INCT-IQ), Grant No. 465469/2014-0.

\bibliography{references}

\begin{thebibliography}{53}%
\makeatletter
\providecommand \@ifxundefined [1]{%
 \@ifx{#1\undefined}
}%
\providecommand \@ifnum [1]{%
 \ifnum #1\expandafter \@firstoftwo
 \else \expandafter \@secondoftwo
 \fi
}%
\providecommand \@ifx [1]{%
 \ifx #1\expandafter \@firstoftwo
 \else \expandafter \@secondoftwo
 \fi
}%
\providecommand \natexlab [1]{#1}%
\providecommand \enquote  [1]{``#1''}%
\providecommand \bibnamefont  [1]{#1}%
\providecommand \bibfnamefont [1]{#1}%
\providecommand \citenamefont [1]{#1}%
\providecommand \href@noop [0]{\@secondoftwo}%
\providecommand \href [0]{\begingroup \@sanitize@url \@href}%
\providecommand \@href[1]{\@@startlink{#1}\@@href}%
\providecommand \@@href[1]{\endgroup#1\@@endlink}%
\providecommand \@sanitize@url [0]{\catcode `\\12\catcode `\$12\catcode `\&12\catcode `\#12\catcode `\^12\catcode `\_12\catcode `\%12\relax}%
\providecommand \@@startlink[1]{}%
\providecommand \@@endlink[0]{}%
\providecommand \url  [0]{\begingroup\@sanitize@url \@url }%
\providecommand \@url [1]{\endgroup\@href {#1}{\urlprefix }}%
\providecommand \urlprefix  [0]{URL }%
\providecommand \Eprint [0]{\href }%
\providecommand \doibase [0]{https://doi.org/}%
\providecommand \selectlanguage [0]{\@gobble}%
\providecommand \bibinfo  [0]{\@secondoftwo}%
\providecommand \bibfield  [0]{\@secondoftwo}%
\providecommand \translation [1]{[#1]}%
\providecommand \BibitemOpen [0]{}%
\providecommand \bibitemStop [0]{}%
\providecommand \bibitemNoStop [0]{.\EOS\space}%
\providecommand \EOS [0]{\spacefactor3000\relax}%
\providecommand \BibitemShut  [1]{\csname bibitem#1\endcsname}%
\let\auto@bib@innerbib\@empty
\bibitem [{\citenamefont {Paszke}\ \emph {et~al.}(2019)\citenamefont {Paszke}, \citenamefont {Gross}, \citenamefont {Massa}, \citenamefont {Lerer}, \citenamefont {Bradbury}, \citenamefont {Chanan}, \citenamefont {Killeen}, \citenamefont {Lin}, \citenamefont {Gimelshein}, \citenamefont {Antiga}, \citenamefont {Desmaison}, \citenamefont {Köpf}, \citenamefont {Yang}, \citenamefont {DeVito}, \citenamefont {Raison}, \citenamefont {Tejani}, \citenamefont {Chilamkurthy}, \citenamefont {Steiner}, \citenamefont {Fang}, \citenamefont {Bai},\ and\ \citenamefont {Chintala}}]{paszke2019pytorch}%
  \BibitemOpen
  \bibfield  {author} {\bibinfo {author} {\bibfnamefont {A.}~\bibnamefont {Paszke}}, \bibinfo {author} {\bibfnamefont {S.}~\bibnamefont {Gross}}, \bibinfo {author} {\bibfnamefont {F.}~\bibnamefont {Massa}}, \bibinfo {author} {\bibfnamefont {A.}~\bibnamefont {Lerer}}, \bibinfo {author} {\bibfnamefont {J.}~\bibnamefont {Bradbury}}, \bibinfo {author} {\bibfnamefont {G.}~\bibnamefont {Chanan}}, \bibinfo {author} {\bibfnamefont {T.}~\bibnamefont {Killeen}}, \bibinfo {author} {\bibfnamefont {Z.}~\bibnamefont {Lin}}, \bibinfo {author} {\bibfnamefont {N.}~\bibnamefont {Gimelshein}}, \bibinfo {author} {\bibfnamefont {L.}~\bibnamefont {Antiga}}, \bibinfo {author} {\bibfnamefont {A.}~\bibnamefont {Desmaison}}, \bibinfo {author} {\bibfnamefont {A.}~\bibnamefont {Köpf}}, \bibinfo {author} {\bibfnamefont {E.}~\bibnamefont {Yang}}, \bibinfo {author} {\bibfnamefont {Z.}~\bibnamefont {DeVito}}, \bibinfo {author} {\bibfnamefont {M.}~\bibnamefont {Raison}}, \bibinfo {author} {\bibfnamefont {A.}~\bibnamefont {Tejani}}, \bibinfo
  {author} {\bibfnamefont {S.}~\bibnamefont {Chilamkurthy}}, \bibinfo {author} {\bibfnamefont {B.}~\bibnamefont {Steiner}}, \bibinfo {author} {\bibfnamefont {L.}~\bibnamefont {Fang}}, \bibinfo {author} {\bibfnamefont {J.}~\bibnamefont {Bai}},\ and\ \bibinfo {author} {\bibfnamefont {S.}~\bibnamefont {Chintala}},\ }\href@noop {} {\bibinfo {title} {Pytorch: An imperative style, high-performance deep learning library}} (\bibinfo {year} {2019}),\ \Eprint {https://arxiv.org/abs/arXiv:1912.01703} {arXiv:1912.01703} \BibitemShut {NoStop}%
\bibitem [{\citenamefont {Bradbury}\ \emph {et~al.}(2018)\citenamefont {Bradbury}, \citenamefont {Frostig}, \citenamefont {Hawkins}, \citenamefont {Johnson}, \citenamefont {Leary}, \citenamefont {Maclaurin}, \citenamefont {Necula}, \citenamefont {Paszke}, \citenamefont {Vander{P}las}, \citenamefont {Wanderman-{M}ilne},\ and\ \citenamefont {Zhang}}]{jax2018github}%
  \BibitemOpen
  \bibfield  {author} {\bibinfo {author} {\bibfnamefont {J.}~\bibnamefont {Bradbury}}, \bibinfo {author} {\bibfnamefont {R.}~\bibnamefont {Frostig}}, \bibinfo {author} {\bibfnamefont {P.}~\bibnamefont {Hawkins}}, \bibinfo {author} {\bibfnamefont {M.~J.}\ \bibnamefont {Johnson}}, \bibinfo {author} {\bibfnamefont {C.}~\bibnamefont {Leary}}, \bibinfo {author} {\bibfnamefont {D.}~\bibnamefont {Maclaurin}}, \bibinfo {author} {\bibfnamefont {G.}~\bibnamefont {Necula}}, \bibinfo {author} {\bibfnamefont {A.}~\bibnamefont {Paszke}}, \bibinfo {author} {\bibfnamefont {J.}~\bibnamefont {Vander{P}las}}, \bibinfo {author} {\bibfnamefont {S.}~\bibnamefont {Wanderman-{M}ilne}},\ and\ \bibinfo {author} {\bibfnamefont {Q.}~\bibnamefont {Zhang}},\ }\href {http://github.com/google/jax} {\bibinfo {title} {{JAX}: composable transformations of {P}ython+{N}um{P}y programs}} (\bibinfo {year} {2018})\BibitemShut {NoStop}%
\bibitem [{\citenamefont {Blondel}\ and\ \citenamefont {Roulet}(2024)}]{Blondel2024}%
  \BibitemOpen
  \bibfield  {author} {\bibinfo {author} {\bibfnamefont {M.}~\bibnamefont {Blondel}}\ and\ \bibinfo {author} {\bibfnamefont {V.}~\bibnamefont {Roulet}},\ }\href {https://doi.org/10.48550/arXiv.2403.14606} {\bibinfo {title} {The elements of differentiable programming}} (\bibinfo {year} {2024}),\ \Eprint {https://arxiv.org/abs/arXiv:2403.14606} {arXiv:2403.14606} \BibitemShut {NoStop}%
\bibitem [{\citenamefont {Goodfellow}\ \emph {et~al.}(2016)\citenamefont {Goodfellow}, \citenamefont {Bengio},\ and\ \citenamefont {Courville}}]{Goodfellow2016}%
  \BibitemOpen
  \bibfield  {author} {\bibinfo {author} {\bibfnamefont {I.}~\bibnamefont {Goodfellow}}, \bibinfo {author} {\bibfnamefont {Y.}~\bibnamefont {Bengio}},\ and\ \bibinfo {author} {\bibfnamefont {A.}~\bibnamefont {Courville}},\ }\href@noop {} {\emph {\bibinfo {title} {Deep Learning}}}\ (\bibinfo  {publisher} {The MIT Press},\ \bibinfo {year} {2016})\BibitemShut {NoStop}%
\bibitem [{\citenamefont {Jouppi}\ \emph {et~al.}(2017)\citenamefont {Jouppi}, \citenamefont {Young}, \citenamefont {Patil}, \citenamefont {Patterson}, \citenamefont {Agrawal}, \citenamefont {Bajwa}, \citenamefont {Bates}, \citenamefont {Bhatia}, \citenamefont {Boden}, \citenamefont {Borchers}, \citenamefont {Boyle}, \citenamefont {luc Cantin}, \citenamefont {Chao}, \citenamefont {Clark}, \citenamefont {Coriell}, \citenamefont {Daley}, \citenamefont {Dau}, \citenamefont {Dean}, \citenamefont {Gelb}, \citenamefont {Ghaemmaghami}, \citenamefont {Gottipati}, \citenamefont {Gulland}, \citenamefont {Hagmann}, \citenamefont {Ho}, \citenamefont {Hogberg}, \citenamefont {Hu}, \citenamefont {Hundt}, \citenamefont {Hurt}, \citenamefont {Ibarz}, \citenamefont {Jaffey}, \citenamefont {Jaworski}, \citenamefont {Kaplan}, \citenamefont {Khaitan}, \citenamefont {Koch}, \citenamefont {Kumar}, \citenamefont {Lacy}, \citenamefont {Laudon}, \citenamefont {Law}, \citenamefont {Le}, \citenamefont {Leary}, \citenamefont {Liu},
  \citenamefont {Lucke}, \citenamefont {Lundin}, \citenamefont {MacKean}, \citenamefont {Maggiore}, \citenamefont {Mahony}, \citenamefont {Miller}, \citenamefont {Nagarajan}, \citenamefont {Narayanaswami}, \citenamefont {Ni}, \citenamefont {Nix}, \citenamefont {Norrie}, \citenamefont {Omernick}, \citenamefont {Penukonda}, \citenamefont {Phelps}, \citenamefont {Ross}, \citenamefont {Ross}, \citenamefont {Salek}, \citenamefont {Samadiani}, \citenamefont {Severn}, \citenamefont {Sizikov}, \citenamefont {Snelham}, \citenamefont {Souter}, \citenamefont {Steinberg}, \citenamefont {Swing}, \citenamefont {Tan}, \citenamefont {Thorson}, \citenamefont {Tian}, \citenamefont {Toma}, \citenamefont {Tuttle}, \citenamefont {Vasudevan}, \citenamefont {Walter}, \citenamefont {Wang}, \citenamefont {Wilcox},\ and\ \citenamefont {Yoon}}]{tpupaper}%
  \BibitemOpen
  \bibfield  {author} {\bibinfo {author} {\bibfnamefont {N.~P.}\ \bibnamefont {Jouppi}}, \bibinfo {author} {\bibfnamefont {C.}~\bibnamefont {Young}}, \bibinfo {author} {\bibfnamefont {N.}~\bibnamefont {Patil}}, \bibinfo {author} {\bibfnamefont {D.}~\bibnamefont {Patterson}}, \bibinfo {author} {\bibfnamefont {G.}~\bibnamefont {Agrawal}}, \bibinfo {author} {\bibfnamefont {R.}~\bibnamefont {Bajwa}}, \bibinfo {author} {\bibfnamefont {S.}~\bibnamefont {Bates}}, \bibinfo {author} {\bibfnamefont {S.}~\bibnamefont {Bhatia}}, \bibinfo {author} {\bibfnamefont {N.}~\bibnamefont {Boden}}, \bibinfo {author} {\bibfnamefont {A.}~\bibnamefont {Borchers}}, \bibinfo {author} {\bibfnamefont {R.}~\bibnamefont {Boyle}}, \bibinfo {author} {\bibfnamefont {P.}~\bibnamefont {luc Cantin}}, \bibinfo {author} {\bibfnamefont {C.}~\bibnamefont {Chao}}, \bibinfo {author} {\bibfnamefont {C.}~\bibnamefont {Clark}}, \bibinfo {author} {\bibfnamefont {J.}~\bibnamefont {Coriell}}, \bibinfo {author} {\bibfnamefont {M.}~\bibnamefont {Daley}},
  \bibinfo {author} {\bibfnamefont {M.}~\bibnamefont {Dau}}, \bibinfo {author} {\bibfnamefont {J.}~\bibnamefont {Dean}}, \bibinfo {author} {\bibfnamefont {B.}~\bibnamefont {Gelb}}, \bibinfo {author} {\bibfnamefont {T.~V.}\ \bibnamefont {Ghaemmaghami}}, \bibinfo {author} {\bibfnamefont {R.}~\bibnamefont {Gottipati}}, \bibinfo {author} {\bibfnamefont {W.}~\bibnamefont {Gulland}}, \bibinfo {author} {\bibfnamefont {R.}~\bibnamefont {Hagmann}}, \bibinfo {author} {\bibfnamefont {C.~R.}\ \bibnamefont {Ho}}, \bibinfo {author} {\bibfnamefont {D.}~\bibnamefont {Hogberg}}, \bibinfo {author} {\bibfnamefont {J.}~\bibnamefont {Hu}}, \bibinfo {author} {\bibfnamefont {R.}~\bibnamefont {Hundt}}, \bibinfo {author} {\bibfnamefont {D.}~\bibnamefont {Hurt}}, \bibinfo {author} {\bibfnamefont {J.}~\bibnamefont {Ibarz}}, \bibinfo {author} {\bibfnamefont {A.}~\bibnamefont {Jaffey}}, \bibinfo {author} {\bibfnamefont {A.}~\bibnamefont {Jaworski}}, \bibinfo {author} {\bibfnamefont {A.}~\bibnamefont {Kaplan}}, \bibinfo {author}
  {\bibfnamefont {H.}~\bibnamefont {Khaitan}}, \bibinfo {author} {\bibfnamefont {A.}~\bibnamefont {Koch}}, \bibinfo {author} {\bibfnamefont {N.}~\bibnamefont {Kumar}}, \bibinfo {author} {\bibfnamefont {S.}~\bibnamefont {Lacy}}, \bibinfo {author} {\bibfnamefont {J.}~\bibnamefont {Laudon}}, \bibinfo {author} {\bibfnamefont {J.}~\bibnamefont {Law}}, \bibinfo {author} {\bibfnamefont {D.}~\bibnamefont {Le}}, \bibinfo {author} {\bibfnamefont {C.}~\bibnamefont {Leary}}, \bibinfo {author} {\bibfnamefont {Z.}~\bibnamefont {Liu}}, \bibinfo {author} {\bibfnamefont {K.}~\bibnamefont {Lucke}}, \bibinfo {author} {\bibfnamefont {A.}~\bibnamefont {Lundin}}, \bibinfo {author} {\bibfnamefont {G.}~\bibnamefont {MacKean}}, \bibinfo {author} {\bibfnamefont {A.}~\bibnamefont {Maggiore}}, \bibinfo {author} {\bibfnamefont {M.}~\bibnamefont {Mahony}}, \bibinfo {author} {\bibfnamefont {K.}~\bibnamefont {Miller}}, \bibinfo {author} {\bibfnamefont {R.}~\bibnamefont {Nagarajan}}, \bibinfo {author} {\bibfnamefont {R.}~\bibnamefont
  {Narayanaswami}}, \bibinfo {author} {\bibfnamefont {R.}~\bibnamefont {Ni}}, \bibinfo {author} {\bibfnamefont {K.}~\bibnamefont {Nix}}, \bibinfo {author} {\bibfnamefont {T.}~\bibnamefont {Norrie}}, \bibinfo {author} {\bibfnamefont {M.}~\bibnamefont {Omernick}}, \bibinfo {author} {\bibfnamefont {N.}~\bibnamefont {Penukonda}}, \bibinfo {author} {\bibfnamefont {A.}~\bibnamefont {Phelps}}, \bibinfo {author} {\bibfnamefont {J.}~\bibnamefont {Ross}}, \bibinfo {author} {\bibfnamefont {M.}~\bibnamefont {Ross}}, \bibinfo {author} {\bibfnamefont {A.}~\bibnamefont {Salek}}, \bibinfo {author} {\bibfnamefont {E.}~\bibnamefont {Samadiani}}, \bibinfo {author} {\bibfnamefont {C.}~\bibnamefont {Severn}}, \bibinfo {author} {\bibfnamefont {G.}~\bibnamefont {Sizikov}}, \bibinfo {author} {\bibfnamefont {M.}~\bibnamefont {Snelham}}, \bibinfo {author} {\bibfnamefont {J.}~\bibnamefont {Souter}}, \bibinfo {author} {\bibfnamefont {D.}~\bibnamefont {Steinberg}}, \bibinfo {author} {\bibfnamefont {A.}~\bibnamefont {Swing}}, \bibinfo
  {author} {\bibfnamefont {M.}~\bibnamefont {Tan}}, \bibinfo {author} {\bibfnamefont {G.}~\bibnamefont {Thorson}}, \bibinfo {author} {\bibfnamefont {B.}~\bibnamefont {Tian}}, \bibinfo {author} {\bibfnamefont {H.}~\bibnamefont {Toma}}, \bibinfo {author} {\bibfnamefont {E.}~\bibnamefont {Tuttle}}, \bibinfo {author} {\bibfnamefont {V.}~\bibnamefont {Vasudevan}}, \bibinfo {author} {\bibfnamefont {R.}~\bibnamefont {Walter}}, \bibinfo {author} {\bibfnamefont {W.}~\bibnamefont {Wang}}, \bibinfo {author} {\bibfnamefont {E.}~\bibnamefont {Wilcox}},\ and\ \bibinfo {author} {\bibfnamefont {D.~H.}\ \bibnamefont {Yoon}},\ }\href@noop {} {\bibinfo {title} {In-datacenter performance analysis of a tensor processing unit}} (\bibinfo {year} {2017}),\ \Eprint {https://arxiv.org/abs/arXiv:1704.04760} {arXiv:arXiv:1704.04760} \BibitemShut {NoStop}%
\bibitem [{\citenamefont {Chae}\ \emph {et~al.}(2024)\citenamefont {Chae}, \citenamefont {Choi},\ and\ \citenamefont {Kim}}]{chae_elementary_2024}%
  \BibitemOpen
  \bibfield  {author} {\bibinfo {author} {\bibfnamefont {E.}~\bibnamefont {Chae}}, \bibinfo {author} {\bibfnamefont {J.}~\bibnamefont {Choi}},\ and\ \bibinfo {author} {\bibfnamefont {J.}~\bibnamefont {Kim}},\ }\bibfield  {title} {\bibinfo {title} {An elementary review on basic principles and developments of qubits for quantum computing},\ }\bibfield  {journal} {\bibinfo  {journal} {Nano Convergence}\ }\textbf {\bibinfo {volume} {11}},\ \href {https://doi.org/10.1186/s40580-024-00418-5} {10.1186/s40580-024-00418-5} (\bibinfo {year} {2024})\BibitemShut {NoStop}%
\bibitem [{\citenamefont {Brylinski}\ and\ \citenamefont {Brylinski}(2001)}]{Brylinski2001UniversalQG}%
  \BibitemOpen
  \bibfield  {author} {\bibinfo {author} {\bibfnamefont {J.-L.}\ \bibnamefont {Brylinski}}\ and\ \bibinfo {author} {\bibfnamefont {R.}~\bibnamefont {Brylinski}},\ }\href {https://arxiv.org/abs/quant-ph/0108062} {\bibinfo {title} {Universal quantum gates}} (\bibinfo {year} {2001}),\ \Eprint {https://arxiv.org/abs/quant-ph/0108062} {arXiv:quant-ph/0108062} \BibitemShut {NoStop}%
\bibitem [{\citenamefont {Wach}\ \emph {et~al.}(2023)\citenamefont {Wach}, \citenamefont {Rudolph}, \citenamefont {Jendrzejewski},\ and\ \citenamefont {Schmitt}}]{Wach_2023}%
  \BibitemOpen
  \bibfield  {author} {\bibinfo {author} {\bibfnamefont {N.~L.}\ \bibnamefont {Wach}}, \bibinfo {author} {\bibfnamefont {M.~S.}\ \bibnamefont {Rudolph}}, \bibinfo {author} {\bibfnamefont {F.}~\bibnamefont {Jendrzejewski}},\ and\ \bibinfo {author} {\bibfnamefont {S.}~\bibnamefont {Schmitt}},\ }\bibfield  {title} {\bibinfo {title} {Data re-uploading with a single qudit},\ }\bibfield  {journal} {\bibinfo  {journal} {Quantum Machine Intelligence}\ }\textbf {\bibinfo {volume} {5}},\ \href {https://doi.org/10.1007/s42484-023-00125-0} {10.1007/s42484-023-00125-0} (\bibinfo {year} {2023})\BibitemShut {NoStop}%
\bibitem [{\citenamefont {Goyal}\ \emph {et~al.}(2014)\citenamefont {Goyal}, \citenamefont {Boukama-Dzoussi}, \citenamefont {Ghosh}, \citenamefont {Roux},\ and\ \citenamefont {Konrad}}]{goyal_qudit-teleportation_2014}%
  \BibitemOpen
  \bibfield  {author} {\bibinfo {author} {\bibfnamefont {S.~K.}\ \bibnamefont {Goyal}}, \bibinfo {author} {\bibfnamefont {P.~E.}\ \bibnamefont {Boukama-Dzoussi}}, \bibinfo {author} {\bibfnamefont {S.}~\bibnamefont {Ghosh}}, \bibinfo {author} {\bibfnamefont {F.~S.}\ \bibnamefont {Roux}},\ and\ \bibinfo {author} {\bibfnamefont {T.}~\bibnamefont {Konrad}},\ }\bibfield  {title} {\bibinfo {title} {Qudit-{Teleportation} for photons with linear optics},\ }\href {https://doi.org/10.1038/srep04543} {\bibfield  {journal} {\bibinfo  {journal} {Scientific Reports}\ }\textbf {\bibinfo {volume} {4}},\ \bibinfo {pages} {4543} (\bibinfo {year} {2014})}\BibitemShut {NoStop}%
\bibitem [{\citenamefont {Wang}\ \emph {et~al.}(2025)\citenamefont {Wang}, \citenamefont {Parker}, \citenamefont {Champion},\ and\ \citenamefont {Blok}}]{qudit_info2}%
  \BibitemOpen
  \bibfield  {author} {\bibinfo {author} {\bibfnamefont {Z.}~\bibnamefont {Wang}}, \bibinfo {author} {\bibfnamefont {R.~W.}\ \bibnamefont {Parker}}, \bibinfo {author} {\bibfnamefont {E.}~\bibnamefont {Champion}},\ and\ \bibinfo {author} {\bibfnamefont {M.~S.}\ \bibnamefont {Blok}},\ }\bibfield  {title} {\bibinfo {title} {High-${E}_{J}/{E}_{C}$ transmon qudits with up to 12 levels},\ }\href {https://doi.org/10.1103/PhysRevApplied.23.034046} {\bibfield  {journal} {\bibinfo  {journal} {Phys. Rev. Appl.}\ }\textbf {\bibinfo {volume} {23}},\ \bibinfo {pages} {034046} (\bibinfo {year} {2025})}\BibitemShut {NoStop}%
\bibitem [{\citenamefont {Ogunkoya}\ \emph {et~al.}(2024)\citenamefont {Ogunkoya}, \citenamefont {Kim}, \citenamefont {Peng}, \citenamefont {\"Ozg\"uler},\ and\ \citenamefont {Alexeev}}]{quditsimulation}%
  \BibitemOpen
  \bibfield  {author} {\bibinfo {author} {\bibfnamefont {O.}~\bibnamefont {Ogunkoya}}, \bibinfo {author} {\bibfnamefont {J.}~\bibnamefont {Kim}}, \bibinfo {author} {\bibfnamefont {B.}~\bibnamefont {Peng}}, \bibinfo {author} {\bibfnamefont {A.~B. i. e. i. f. m.~c.}\ \bibnamefont {\"Ozg\"uler}},\ and\ \bibinfo {author} {\bibfnamefont {Y.}~\bibnamefont {Alexeev}},\ }\bibfield  {title} {\bibinfo {title} {Qutrit circuits and algebraic relations: A pathway to efficient spin-1 hamiltonian simulation},\ }\href {https://doi.org/10.1103/PhysRevA.109.012426} {\bibfield  {journal} {\bibinfo  {journal} {Phys. Rev. A}\ }\textbf {\bibinfo {volume} {109}},\ \bibinfo {pages} {012426} (\bibinfo {year} {2024})}\BibitemShut {NoStop}%
\bibitem [{\citenamefont {Tacchino}\ \emph {et~al.}(2021)\citenamefont {Tacchino}, \citenamefont {Chiesa}, \citenamefont {Sessoli}, \citenamefont {Tavernelli},\ and\ \citenamefont {Carretta}}]{tacchino_proposal_2021}%
  \BibitemOpen
  \bibfield  {author} {\bibinfo {author} {\bibfnamefont {F.}~\bibnamefont {Tacchino}}, \bibinfo {author} {\bibfnamefont {A.}~\bibnamefont {Chiesa}}, \bibinfo {author} {\bibfnamefont {R.}~\bibnamefont {Sessoli}}, \bibinfo {author} {\bibfnamefont {I.}~\bibnamefont {Tavernelli}},\ and\ \bibinfo {author} {\bibfnamefont {S.}~\bibnamefont {Carretta}},\ }\bibfield  {title} {\bibinfo {title} {A proposal for using molecular spin qudits as quantum simulators of light-matter interactions},\ }\href {https://doi.org/10.1039/D1TC00851J} {\bibfield  {journal} {\bibinfo  {journal} {Journal of Materials Chemistry C}\ }\textbf {\bibinfo {volume} {9}},\ \bibinfo {pages} {10266} (\bibinfo {year} {2021})}\BibitemShut {NoStop}%
\bibitem [{\citenamefont {Bourennane}\ \emph {et~al.}(2001)\citenamefont {Bourennane}, \citenamefont {Karlsson},\ and\ \citenamefont {Bj\"ork}}]{qudit_crypt}%
  \BibitemOpen
  \bibfield  {author} {\bibinfo {author} {\bibfnamefont {M.}~\bibnamefont {Bourennane}}, \bibinfo {author} {\bibfnamefont {A.}~\bibnamefont {Karlsson}},\ and\ \bibinfo {author} {\bibfnamefont {G.}~\bibnamefont {Bj\"ork}},\ }\bibfield  {title} {\bibinfo {title} {Quantum key distribution using multilevel encoding},\ }\href {https://doi.org/10.1103/PhysRevA.64.012306} {\bibfield  {journal} {\bibinfo  {journal} {Phys. Rev. A}\ }\textbf {\bibinfo {volume} {64}},\ \bibinfo {pages} {012306} (\bibinfo {year} {2001})}\BibitemShut {NoStop}%
\bibitem [{\citenamefont {Litteken}\ \emph {et~al.}(2022)\citenamefont {Litteken}, \citenamefont {Baker},\ and\ \citenamefont {Chong}}]{quditcomm}%
  \BibitemOpen
  \bibfield  {author} {\bibinfo {author} {\bibfnamefont {A.}~\bibnamefont {Litteken}}, \bibinfo {author} {\bibfnamefont {J.~M.}\ \bibnamefont {Baker}},\ and\ \bibinfo {author} {\bibfnamefont {F.~T.}\ \bibnamefont {Chong}},\ }\bibfield  {title} {\bibinfo {title} {Communication trade offs in intermediate qudit circuits},\ }in\ \href {https://doi.org/10.1109/ISMVL52857.2022.00014} {\emph {\bibinfo {booktitle} {2022 IEEE 52nd International Symposium on Multiple-Valued Logic (ISMVL)}}}\ (\bibinfo {year} {2022})\ pp.\ \bibinfo {pages} {43--49}\BibitemShut {NoStop}%
\bibitem [{\citenamefont {Chizzini}\ \emph {et~al.}(2022)\citenamefont {Chizzini}, \citenamefont {Crippa}, \citenamefont {Zaccardi}, \citenamefont {Macaluso}, \citenamefont {Carretta}, \citenamefont {Chiesa},\ and\ \citenamefont {Santini}}]{chizzini_quantum_2022}%
  \BibitemOpen
  \bibfield  {author} {\bibinfo {author} {\bibfnamefont {M.}~\bibnamefont {Chizzini}}, \bibinfo {author} {\bibfnamefont {L.}~\bibnamefont {Crippa}}, \bibinfo {author} {\bibfnamefont {L.}~\bibnamefont {Zaccardi}}, \bibinfo {author} {\bibfnamefont {E.}~\bibnamefont {Macaluso}}, \bibinfo {author} {\bibfnamefont {S.}~\bibnamefont {Carretta}}, \bibinfo {author} {\bibfnamefont {A.}~\bibnamefont {Chiesa}},\ and\ \bibinfo {author} {\bibfnamefont {P.}~\bibnamefont {Santini}},\ }\bibfield  {title} {\bibinfo {title} {Quantum error correction with molecular spin qudits},\ }\href {https://doi.org/10.1039/D2CP01228F} {\bibfield  {journal} {\bibinfo  {journal} {Physical Chemistry Chemical Physics}\ }\textbf {\bibinfo {volume} {24}},\ \bibinfo {pages} {20030} (\bibinfo {year} {2022})}\BibitemShut {NoStop}%
\bibitem [{\citenamefont {Tripathi}\ \emph {et~al.}(2025)\citenamefont {Tripathi}, \citenamefont {Goss}, \citenamefont {Vezvaee}, \citenamefont {Nguyen}, \citenamefont {Siddiqi},\ and\ \citenamefont {Lidar}}]{qudit_error2}%
  \BibitemOpen
  \bibfield  {author} {\bibinfo {author} {\bibfnamefont {V.}~\bibnamefont {Tripathi}}, \bibinfo {author} {\bibfnamefont {N.}~\bibnamefont {Goss}}, \bibinfo {author} {\bibfnamefont {A.}~\bibnamefont {Vezvaee}}, \bibinfo {author} {\bibfnamefont {L.~B.}\ \bibnamefont {Nguyen}}, \bibinfo {author} {\bibfnamefont {I.}~\bibnamefont {Siddiqi}},\ and\ \bibinfo {author} {\bibfnamefont {D.~A.}\ \bibnamefont {Lidar}},\ }\bibfield  {title} {\bibinfo {title} {Qudit dynamical decoupling on a superconducting quantum processor},\ }\href {https://doi.org/10.1103/PhysRevLett.134.050601} {\bibfield  {journal} {\bibinfo  {journal} {Phys. Rev. Lett.}\ }\textbf {\bibinfo {volume} {134}},\ \bibinfo {pages} {050601} (\bibinfo {year} {2025})}\BibitemShut {NoStop}%
\bibitem [{\citenamefont {Caro}\ \emph {et~al.}(2022)\citenamefont {Caro}, \citenamefont {Huang}, \citenamefont {Cerezo}, \citenamefont {Sharma}, \citenamefont {Sornborger}, \citenamefont {Cincio},\ and\ \citenamefont {Coles}}]{caro_generalization_2022}%
  \BibitemOpen
  \bibfield  {author} {\bibinfo {author} {\bibfnamefont {M.~C.}\ \bibnamefont {Caro}}, \bibinfo {author} {\bibfnamefont {H.-Y.}\ \bibnamefont {Huang}}, \bibinfo {author} {\bibfnamefont {M.}~\bibnamefont {Cerezo}}, \bibinfo {author} {\bibfnamefont {K.}~\bibnamefont {Sharma}}, \bibinfo {author} {\bibfnamefont {A.}~\bibnamefont {Sornborger}}, \bibinfo {author} {\bibfnamefont {L.}~\bibnamefont {Cincio}},\ and\ \bibinfo {author} {\bibfnamefont {P.~J.}\ \bibnamefont {Coles}},\ }\bibfield  {title} {\bibinfo {title} {Generalization in quantum machine learning from few training data},\ }\href {https://doi.org/10.1038/s41467-022-32550-3} {\bibfield  {journal} {\bibinfo  {journal} {Nature Communications}\ }\textbf {\bibinfo {volume} {13}},\ \bibinfo {pages} {4919} (\bibinfo {year} {2022})}\BibitemShut {NoStop}%
\bibitem [{\citenamefont {Ding}\ \emph {et~al.}(2024)\citenamefont {Ding}, \citenamefont {Ban}, \citenamefont {Sanz}, \citenamefont {Martín-Guerrero},\ and\ \citenamefont {Chen}}]{ding2024quantumactivelearning}%
  \BibitemOpen
  \bibfield  {author} {\bibinfo {author} {\bibfnamefont {Y.}~\bibnamefont {Ding}}, \bibinfo {author} {\bibfnamefont {Y.}~\bibnamefont {Ban}}, \bibinfo {author} {\bibfnamefont {M.}~\bibnamefont {Sanz}}, \bibinfo {author} {\bibfnamefont {J.~D.}\ \bibnamefont {Martín-Guerrero}},\ and\ \bibinfo {author} {\bibfnamefont {X.}~\bibnamefont {Chen}},\ }\bibfield  {title} {\bibinfo {title} {Quantum active learning},\ }\href@noop {} {\bibfield  {journal} {\bibinfo  {journal} {arXiv:2405.18230}\ } (\bibinfo {year} {2024})}\BibitemShut {NoStop}%
\bibitem [{\citenamefont {Liu}\ \emph {et~al.}(2021)\citenamefont {Liu}, \citenamefont {Arunachalam},\ and\ \citenamefont {Temme}}]{liu_rigorous_2021}%
  \BibitemOpen
  \bibfield  {author} {\bibinfo {author} {\bibfnamefont {Y.}~\bibnamefont {Liu}}, \bibinfo {author} {\bibfnamefont {S.}~\bibnamefont {Arunachalam}},\ and\ \bibinfo {author} {\bibfnamefont {K.}~\bibnamefont {Temme}},\ }\bibfield  {title} {\bibinfo {title} {A rigorous and robust quantum speed-up in supervised machine learning},\ }\href {https://doi.org/10.1038/s41567-021-01287-z} {\bibfield  {journal} {\bibinfo  {journal} {Nature Physics}\ }\textbf {\bibinfo {volume} {17}},\ \bibinfo {pages} {1013} (\bibinfo {year} {2021})}\BibitemShut {NoStop}%
\bibitem [{\citenamefont {Liao}\ \emph {et~al.}(2024)\citenamefont {Liao}, \citenamefont {Sanders},\ and\ \citenamefont {Byrnes}}]{liao_quadratic_2024}%
  \BibitemOpen
  \bibfield  {author} {\bibinfo {author} {\bibfnamefont {P.}~\bibnamefont {Liao}}, \bibinfo {author} {\bibfnamefont {B.~C.}\ \bibnamefont {Sanders}},\ and\ \bibinfo {author} {\bibfnamefont {T.}~\bibnamefont {Byrnes}},\ }\bibfield  {title} {\bibinfo {title} {Quadratic quantum speedup for perceptron training},\ }\href {https://doi.org/10.1103/PhysRevA.110.062412} {\bibfield  {journal} {\bibinfo  {journal} {Physical Review A}\ }\textbf {\bibinfo {volume} {110}},\ \bibinfo {pages} {062412} (\bibinfo {year} {2024})}\BibitemShut {NoStop}%
\bibitem [{\citenamefont {Cerezo}\ \emph {et~al.}(2022)\citenamefont {Cerezo}, \citenamefont {Verdon}, \citenamefont {Huang}, \citenamefont {Cincio},\ and\ \citenamefont {Coles}}]{cerezo_challenges_2022}%
  \BibitemOpen
  \bibfield  {author} {\bibinfo {author} {\bibfnamefont {M.}~\bibnamefont {Cerezo}}, \bibinfo {author} {\bibfnamefont {G.}~\bibnamefont {Verdon}}, \bibinfo {author} {\bibfnamefont {H.-Y.}\ \bibnamefont {Huang}}, \bibinfo {author} {\bibfnamefont {L.}~\bibnamefont {Cincio}},\ and\ \bibinfo {author} {\bibfnamefont {P.~J.}\ \bibnamefont {Coles}},\ }\bibfield  {title} {\bibinfo {title} {Challenges and opportunities in quantum machine learning},\ }\href {https://doi.org/10.1038/s43588-022-00311-3} {\bibfield  {journal} {\bibinfo  {journal} {Nature Computational Science}\ }\textbf {\bibinfo {volume} {2}},\ \bibinfo {pages} {567} (\bibinfo {year} {2022})}\BibitemShut {NoStop}%
\bibitem [{\citenamefont {Graña}\ \emph {et~al.}(2025)\citenamefont {Graña}, \citenamefont {Varsamopoulos}, \citenamefont {Ando}, \citenamefont {Maeshima},\ and\ \citenamefont {Matsuzawa}}]{grana_materials_2025}%
  \BibitemOpen
  \bibfield  {author} {\bibinfo {author} {\bibfnamefont {I.~F.}\ \bibnamefont {Graña}}, \bibinfo {author} {\bibfnamefont {S.}~\bibnamefont {Varsamopoulos}}, \bibinfo {author} {\bibfnamefont {T.}~\bibnamefont {Ando}}, \bibinfo {author} {\bibfnamefont {H.}~\bibnamefont {Maeshima}},\ and\ \bibinfo {author} {\bibfnamefont {N.~N.}\ \bibnamefont {Matsuzawa}},\ }\href {https://doi.org/10.48550/arXiv.2503.09517} {\bibinfo {title} {Materials {Discovery} {With} {Quantum}-{Enhanced} {Machine} {Learning} {Algorithms}}} (\bibinfo {year} {2025}),\ \bibinfo {note} {arXiv:2503.09517}\BibitemShut {NoStop}%
\bibitem [{\citenamefont {Sajjan}\ \emph {et~al.}(2022)\citenamefont {Sajjan}, \citenamefont {Li}, \citenamefont {Selvarajan}, \citenamefont {Sureshbabu}, \citenamefont {Kale}, \citenamefont {Gupta}, \citenamefont {Singh},\ and\ \citenamefont {Kais}}]{sajjan_quantum_2022}%
  \BibitemOpen
  \bibfield  {author} {\bibinfo {author} {\bibfnamefont {M.}~\bibnamefont {Sajjan}}, \bibinfo {author} {\bibfnamefont {J.}~\bibnamefont {Li}}, \bibinfo {author} {\bibfnamefont {R.}~\bibnamefont {Selvarajan}}, \bibinfo {author} {\bibfnamefont {S.~H.}\ \bibnamefont {Sureshbabu}}, \bibinfo {author} {\bibfnamefont {S.~S.}\ \bibnamefont {Kale}}, \bibinfo {author} {\bibfnamefont {R.}~\bibnamefont {Gupta}}, \bibinfo {author} {\bibfnamefont {V.}~\bibnamefont {Singh}},\ and\ \bibinfo {author} {\bibfnamefont {S.}~\bibnamefont {Kais}},\ }\href {https://doi.org/10.48550/arXiv.2111.00851} {\bibinfo {title} {Quantum {Machine} {Learning} for {Chemistry} and {Physics}}} (\bibinfo {year} {2022}),\ \bibinfo {note} {arXiv:2111.00851}\BibitemShut {NoStop}%
\bibitem [{\citenamefont {Scott}\ and\ \citenamefont {T{\r{u}}ma}(2023)}]{Scott2023}%
  \BibitemOpen
  \bibfield  {author} {\bibinfo {author} {\bibfnamefont {J.}~\bibnamefont {Scott}}\ and\ \bibinfo {author} {\bibfnamefont {M.}~\bibnamefont {T{\r{u}}ma}},\ }\bibinfo {title} {An introduction to sparse matrices},\ in\ \href {https://doi.org/10.1007/978-3-031-25820-6_1} {\emph {\bibinfo {booktitle} {Algorithms for Sparse Linear Systems}}}\ (\bibinfo  {publisher} {Springer International Publishing},\ \bibinfo {address} {Cham},\ \bibinfo {year} {2023})\ pp.\ \bibinfo {pages} {1--18}\BibitemShut {NoStop}%
\bibitem [{\citenamefont {Deller}\ \emph {et~al.}(2023)\citenamefont {Deller}, \citenamefont {Schmitt}, \citenamefont {Lewenstein}, \citenamefont {Lenk}, \citenamefont {Federer}, \citenamefont {Jendrzejewski}, \citenamefont {Hauke},\ and\ \citenamefont {Kasper}}]{Deller_2023}%
  \BibitemOpen
  \bibfield  {author} {\bibinfo {author} {\bibfnamefont {Y.}~\bibnamefont {Deller}}, \bibinfo {author} {\bibfnamefont {S.}~\bibnamefont {Schmitt}}, \bibinfo {author} {\bibfnamefont {M.}~\bibnamefont {Lewenstein}}, \bibinfo {author} {\bibfnamefont {S.}~\bibnamefont {Lenk}}, \bibinfo {author} {\bibfnamefont {M.}~\bibnamefont {Federer}}, \bibinfo {author} {\bibfnamefont {F.}~\bibnamefont {Jendrzejewski}}, \bibinfo {author} {\bibfnamefont {P.}~\bibnamefont {Hauke}},\ and\ \bibinfo {author} {\bibfnamefont {V.}~\bibnamefont {Kasper}},\ }\bibfield  {title} {\bibinfo {title} {Quantum approximate optimization algorithm for qudit systems},\ }\bibfield  {journal} {\bibinfo  {journal} {Physical Review A}\ }\textbf {\bibinfo {volume} {107}},\ \href {https://doi.org/10.1103/physreva.107.062410} {10.1103/physreva.107.062410} (\bibinfo {year} {2023})\BibitemShut {NoStop}%
\bibitem [{\citenamefont {Alchieri}\ \emph {et~al.}(2021)\citenamefont {Alchieri}, \citenamefont {Badalotti}, \citenamefont {Bonardi},\ and\ \citenamefont {Bianco}}]{alchieri_introduction_2021}%
  \BibitemOpen
  \bibfield  {author} {\bibinfo {author} {\bibfnamefont {L.}~\bibnamefont {Alchieri}}, \bibinfo {author} {\bibfnamefont {D.}~\bibnamefont {Badalotti}}, \bibinfo {author} {\bibfnamefont {P.}~\bibnamefont {Bonardi}},\ and\ \bibinfo {author} {\bibfnamefont {S.}~\bibnamefont {Bianco}},\ }\bibfield  {title} {\bibinfo {title} {An introduction to quantum machine learning: from quantum logic to quantum deep learning},\ }\href {https://doi.org/10.1007/s42484-021-00056-8} {\bibfield  {journal} {\bibinfo  {journal} {Quantum Machine Intelligence}\ }\textbf {\bibinfo {volume} {3}},\ \bibinfo {pages} {28} (\bibinfo {year} {2021})}\BibitemShut {NoStop}%
\bibitem [{\citenamefont {Cerezo}\ \emph {et~al.}(2021{\natexlab{a}})\citenamefont {Cerezo}, \citenamefont {Arrasmith}, \citenamefont {Babbush}, \citenamefont {Benjamin}, \citenamefont {Endo}, \citenamefont {Fujii}, \citenamefont {McClean}, \citenamefont {Mitarai}, \citenamefont {Yuan}, \citenamefont {Cincio},\ and\ \citenamefont {Coles}}]{cerezo_variational_2021}%
  \BibitemOpen
  \bibfield  {author} {\bibinfo {author} {\bibfnamefont {M.}~\bibnamefont {Cerezo}}, \bibinfo {author} {\bibfnamefont {A.}~\bibnamefont {Arrasmith}}, \bibinfo {author} {\bibfnamefont {R.}~\bibnamefont {Babbush}}, \bibinfo {author} {\bibfnamefont {S.~C.}\ \bibnamefont {Benjamin}}, \bibinfo {author} {\bibfnamefont {S.}~\bibnamefont {Endo}}, \bibinfo {author} {\bibfnamefont {K.}~\bibnamefont {Fujii}}, \bibinfo {author} {\bibfnamefont {J.~R.}\ \bibnamefont {McClean}}, \bibinfo {author} {\bibfnamefont {K.}~\bibnamefont {Mitarai}}, \bibinfo {author} {\bibfnamefont {X.}~\bibnamefont {Yuan}}, \bibinfo {author} {\bibfnamefont {L.}~\bibnamefont {Cincio}},\ and\ \bibinfo {author} {\bibfnamefont {P.~J.}\ \bibnamefont {Coles}},\ }\bibfield  {title} {\bibinfo {title} {Variational quantum algorithms},\ }\href {https://doi.org/10.1038/s42254-021-00348-9} {\bibfield  {journal} {\bibinfo  {journal} {Nature Reviews Physics}\ }\textbf {\bibinfo {volume} {3}},\ \bibinfo {pages} {625} (\bibinfo {year}
  {2021}{\natexlab{a}})}\BibitemShut {NoStop}%
\bibitem [{\citenamefont {Roca-Jerat}\ \emph {et~al.}(2023)\citenamefont {Roca-Jerat}, \citenamefont {Román-Roche},\ and\ \citenamefont {Zueco}}]{rocajerat2023qudit}%
  \BibitemOpen
  \bibfield  {author} {\bibinfo {author} {\bibfnamefont {S.}~\bibnamefont {Roca-Jerat}}, \bibinfo {author} {\bibfnamefont {J.}~\bibnamefont {Román-Roche}},\ and\ \bibinfo {author} {\bibfnamefont {D.}~\bibnamefont {Zueco}},\ }\href@noop {} {\bibinfo {title} {Qudit machine learning}} (\bibinfo {year} {2023}),\ \Eprint {https://arxiv.org/abs/2308.16230} {arXiv:2308.16230} \BibitemShut {NoStop}%
\bibitem [{\citenamefont {Souza}\ and\ \citenamefont {Portugal}(2025)}]{qudit_ml2}%
  \BibitemOpen
  \bibfield  {author} {\bibinfo {author} {\bibfnamefont {L.~C.}\ \bibnamefont {Souza}}\ and\ \bibinfo {author} {\bibfnamefont {R.}~\bibnamefont {Portugal}},\ }\href {https://arxiv.org/abs/2503.09269} {\bibinfo {title} {Single-qudit quantum neural networks for multiclass classification}} (\bibinfo {year} {2025}),\ \Eprint {https://arxiv.org/abs/2503.09269} {arXiv:2503.09269} \BibitemShut {NoStop}%
\bibitem [{\citenamefont {Fischer}\ \emph {et~al.}(2023)\citenamefont {Fischer}, \citenamefont {Chiesa}, \citenamefont {Tacchino}, \citenamefont {Egger}, \citenamefont {Carretta},\ and\ \citenamefont {Tavernelli}}]{Fischer_2023}%
  \BibitemOpen
  \bibfield  {author} {\bibinfo {author} {\bibfnamefont {L.~E.}\ \bibnamefont {Fischer}}, \bibinfo {author} {\bibfnamefont {A.}~\bibnamefont {Chiesa}}, \bibinfo {author} {\bibfnamefont {F.}~\bibnamefont {Tacchino}}, \bibinfo {author} {\bibfnamefont {D.~J.}\ \bibnamefont {Egger}}, \bibinfo {author} {\bibfnamefont {S.}~\bibnamefont {Carretta}},\ and\ \bibinfo {author} {\bibfnamefont {I.}~\bibnamefont {Tavernelli}},\ }\bibfield  {title} {\bibinfo {title} {Universal qudit gate synthesis for transmons},\ }\bibfield  {journal} {\bibinfo  {journal} {PRX Quantum}\ }\textbf {\bibinfo {volume} {4}},\ \href {https://doi.org/10.1103/prxquantum.4.030327} {10.1103/prxquantum.4.030327} (\bibinfo {year} {2023})\BibitemShut {NoStop}%
\bibitem [{\citenamefont {Kirk}(2007)}]{cudapaper}%
  \BibitemOpen
  \bibfield  {author} {\bibinfo {author} {\bibfnamefont {D.}~\bibnamefont {Kirk}},\ }\bibfield  {title} {\bibinfo {title} {Nvidia cuda software and gpu parallel computing architecture},\ }in\ \href {https://doi.org/10.1145/1296907.1296909} {\emph {\bibinfo {booktitle} {Proceedings of the 6th International Symposium on Memory Management}}},\ \bibinfo {series and number} {ISMM '07}\ (\bibinfo  {publisher} {Association for Computing Machinery},\ \bibinfo {address} {New York, NY, USA},\ \bibinfo {year} {2007})\ p.\ \bibinfo {pages} {103–104}\BibitemShut {NoStop}%
\bibitem [{\citenamefont {Javadi-Abhari}\ \emph {et~al.}(2024)\citenamefont {Javadi-Abhari}, \citenamefont {Treinish}, \citenamefont {Krsulich}, \citenamefont {Wood}, \citenamefont {Lishman}, \citenamefont {Gacon}, \citenamefont {Martiel}, \citenamefont {Nation}, \citenamefont {Bishop}, \citenamefont {Cross}, \citenamefont {Johnson},\ and\ \citenamefont {Gambetta}}]{qiskit2024}%
  \BibitemOpen
  \bibfield  {author} {\bibinfo {author} {\bibfnamefont {A.}~\bibnamefont {Javadi-Abhari}}, \bibinfo {author} {\bibfnamefont {M.}~\bibnamefont {Treinish}}, \bibinfo {author} {\bibfnamefont {K.}~\bibnamefont {Krsulich}}, \bibinfo {author} {\bibfnamefont {C.~J.}\ \bibnamefont {Wood}}, \bibinfo {author} {\bibfnamefont {J.}~\bibnamefont {Lishman}}, \bibinfo {author} {\bibfnamefont {J.}~\bibnamefont {Gacon}}, \bibinfo {author} {\bibfnamefont {S.}~\bibnamefont {Martiel}}, \bibinfo {author} {\bibfnamefont {P.~D.}\ \bibnamefont {Nation}}, \bibinfo {author} {\bibfnamefont {L.~S.}\ \bibnamefont {Bishop}}, \bibinfo {author} {\bibfnamefont {A.~W.}\ \bibnamefont {Cross}}, \bibinfo {author} {\bibfnamefont {B.~R.}\ \bibnamefont {Johnson}},\ and\ \bibinfo {author} {\bibfnamefont {J.~M.}\ \bibnamefont {Gambetta}},\ }\href {https://doi.org/10.48550/arXiv.2405.08810} {\bibinfo {title} {Quantum computing with {Q}iskit}} (\bibinfo {year} {2024}),\ \Eprint {https://arxiv.org/abs/2405.08810} {arXiv:2405.08810} \BibitemShut
  {NoStop}%
\bibitem [{\citenamefont {Bergholm}\ \emph {et~al.}(2022)\citenamefont {Bergholm}, \citenamefont {Izaac}, \citenamefont {Schuld}, \citenamefont {Gogolin}, \citenamefont {Ahmed}, \citenamefont {Ajith}, \citenamefont {Alam}, \citenamefont {Alonso-Linaje}, \citenamefont {AkashNarayanan}, \citenamefont {Asadi}, \citenamefont {Arrazola}, \citenamefont {Azad}, \citenamefont {Banning}, \citenamefont {Blank}, \citenamefont {Bromley}, \citenamefont {Cordier}, \citenamefont {Ceroni}, \citenamefont {Delgado}, \citenamefont {Matteo}, \citenamefont {Dusko}, \citenamefont {Garg}, \citenamefont {Guala}, \citenamefont {Hayes}, \citenamefont {Hill}, \citenamefont {Ijaz}, \citenamefont {Isacsson}, \citenamefont {Ittah}, \citenamefont {Jahangiri}, \citenamefont {Jain}, \citenamefont {Jiang}, \citenamefont {Khandelwal}, \citenamefont {Kottmann}, \citenamefont {Lang}, \citenamefont {Lee}, \citenamefont {Loke}, \citenamefont {Lowe}, \citenamefont {McKiernan}, \citenamefont {Meyer}, \citenamefont {Montañez-Barrera}, \citenamefont
  {Moyard}, \citenamefont {Niu}, \citenamefont {O'Riordan}, \citenamefont {Oud}, \citenamefont {Panigrahi}, \citenamefont {Park}, \citenamefont {Polatajko}, \citenamefont {Quesada}, \citenamefont {Roberts}, \citenamefont {Sá}, \citenamefont {Schoch}, \citenamefont {Shi}, \citenamefont {Shu}, \citenamefont {Sim}, \citenamefont {Singh}, \citenamefont {Strandberg}, \citenamefont {Soni}, \citenamefont {Száva}, \citenamefont {Thabet}, \citenamefont {Vargas-Hernández}, \citenamefont {Vincent}, \citenamefont {Vitucci}, \citenamefont {Weber}, \citenamefont {Wierichs}, \citenamefont {Wiersema}, \citenamefont {Willmann}, \citenamefont {Wong}, \citenamefont {Zhang},\ and\ \citenamefont {Killoran}}]{bergholm2022pennylane}%
  \BibitemOpen
  \bibfield  {author} {\bibinfo {author} {\bibfnamefont {V.}~\bibnamefont {Bergholm}}, \bibinfo {author} {\bibfnamefont {J.}~\bibnamefont {Izaac}}, \bibinfo {author} {\bibfnamefont {M.}~\bibnamefont {Schuld}}, \bibinfo {author} {\bibfnamefont {C.}~\bibnamefont {Gogolin}}, \bibinfo {author} {\bibfnamefont {S.}~\bibnamefont {Ahmed}}, \bibinfo {author} {\bibfnamefont {V.}~\bibnamefont {Ajith}}, \bibinfo {author} {\bibfnamefont {M.~S.}\ \bibnamefont {Alam}}, \bibinfo {author} {\bibfnamefont {G.}~\bibnamefont {Alonso-Linaje}}, \bibinfo {author} {\bibfnamefont {B.}~\bibnamefont {AkashNarayanan}}, \bibinfo {author} {\bibfnamefont {A.}~\bibnamefont {Asadi}}, \bibinfo {author} {\bibfnamefont {J.~M.}\ \bibnamefont {Arrazola}}, \bibinfo {author} {\bibfnamefont {U.}~\bibnamefont {Azad}}, \bibinfo {author} {\bibfnamefont {S.}~\bibnamefont {Banning}}, \bibinfo {author} {\bibfnamefont {C.}~\bibnamefont {Blank}}, \bibinfo {author} {\bibfnamefont {T.~R.}\ \bibnamefont {Bromley}}, \bibinfo {author} {\bibfnamefont {B.~A.}\
  \bibnamefont {Cordier}}, \bibinfo {author} {\bibfnamefont {J.}~\bibnamefont {Ceroni}}, \bibinfo {author} {\bibfnamefont {A.}~\bibnamefont {Delgado}}, \bibinfo {author} {\bibfnamefont {O.~D.}\ \bibnamefont {Matteo}}, \bibinfo {author} {\bibfnamefont {A.}~\bibnamefont {Dusko}}, \bibinfo {author} {\bibfnamefont {T.}~\bibnamefont {Garg}}, \bibinfo {author} {\bibfnamefont {D.}~\bibnamefont {Guala}}, \bibinfo {author} {\bibfnamefont {A.}~\bibnamefont {Hayes}}, \bibinfo {author} {\bibfnamefont {R.}~\bibnamefont {Hill}}, \bibinfo {author} {\bibfnamefont {A.}~\bibnamefont {Ijaz}}, \bibinfo {author} {\bibfnamefont {T.}~\bibnamefont {Isacsson}}, \bibinfo {author} {\bibfnamefont {D.}~\bibnamefont {Ittah}}, \bibinfo {author} {\bibfnamefont {S.}~\bibnamefont {Jahangiri}}, \bibinfo {author} {\bibfnamefont {P.}~\bibnamefont {Jain}}, \bibinfo {author} {\bibfnamefont {E.}~\bibnamefont {Jiang}}, \bibinfo {author} {\bibfnamefont {A.}~\bibnamefont {Khandelwal}}, \bibinfo {author} {\bibfnamefont {K.}~\bibnamefont {Kottmann}},
  \bibinfo {author} {\bibfnamefont {R.~A.}\ \bibnamefont {Lang}}, \bibinfo {author} {\bibfnamefont {C.}~\bibnamefont {Lee}}, \bibinfo {author} {\bibfnamefont {T.}~\bibnamefont {Loke}}, \bibinfo {author} {\bibfnamefont {A.}~\bibnamefont {Lowe}}, \bibinfo {author} {\bibfnamefont {K.}~\bibnamefont {McKiernan}}, \bibinfo {author} {\bibfnamefont {J.~J.}\ \bibnamefont {Meyer}}, \bibinfo {author} {\bibfnamefont {J.~A.}\ \bibnamefont {Montañez-Barrera}}, \bibinfo {author} {\bibfnamefont {R.}~\bibnamefont {Moyard}}, \bibinfo {author} {\bibfnamefont {Z.}~\bibnamefont {Niu}}, \bibinfo {author} {\bibfnamefont {L.~J.}\ \bibnamefont {O'Riordan}}, \bibinfo {author} {\bibfnamefont {S.}~\bibnamefont {Oud}}, \bibinfo {author} {\bibfnamefont {A.}~\bibnamefont {Panigrahi}}, \bibinfo {author} {\bibfnamefont {C.-Y.}\ \bibnamefont {Park}}, \bibinfo {author} {\bibfnamefont {D.}~\bibnamefont {Polatajko}}, \bibinfo {author} {\bibfnamefont {N.}~\bibnamefont {Quesada}}, \bibinfo {author} {\bibfnamefont {C.}~\bibnamefont {Roberts}},
  \bibinfo {author} {\bibfnamefont {N.}~\bibnamefont {Sá}}, \bibinfo {author} {\bibfnamefont {I.}~\bibnamefont {Schoch}}, \bibinfo {author} {\bibfnamefont {B.}~\bibnamefont {Shi}}, \bibinfo {author} {\bibfnamefont {S.}~\bibnamefont {Shu}}, \bibinfo {author} {\bibfnamefont {S.}~\bibnamefont {Sim}}, \bibinfo {author} {\bibfnamefont {A.}~\bibnamefont {Singh}}, \bibinfo {author} {\bibfnamefont {I.}~\bibnamefont {Strandberg}}, \bibinfo {author} {\bibfnamefont {J.}~\bibnamefont {Soni}}, \bibinfo {author} {\bibfnamefont {A.}~\bibnamefont {Száva}}, \bibinfo {author} {\bibfnamefont {S.}~\bibnamefont {Thabet}}, \bibinfo {author} {\bibfnamefont {R.~A.}\ \bibnamefont {Vargas-Hernández}}, \bibinfo {author} {\bibfnamefont {T.}~\bibnamefont {Vincent}}, \bibinfo {author} {\bibfnamefont {N.}~\bibnamefont {Vitucci}}, \bibinfo {author} {\bibfnamefont {M.}~\bibnamefont {Weber}}, \bibinfo {author} {\bibfnamefont {D.}~\bibnamefont {Wierichs}}, \bibinfo {author} {\bibfnamefont {R.}~\bibnamefont {Wiersema}}, \bibinfo {author}
  {\bibfnamefont {M.}~\bibnamefont {Willmann}}, \bibinfo {author} {\bibfnamefont {V.}~\bibnamefont {Wong}}, \bibinfo {author} {\bibfnamefont {S.}~\bibnamefont {Zhang}},\ and\ \bibinfo {author} {\bibfnamefont {N.}~\bibnamefont {Killoran}},\ }\href@noop {} {\bibinfo {title} {Pennylane: Automatic differentiation of hybrid quantum-classical computations}} (\bibinfo {year} {2022}),\ \Eprint {https://arxiv.org/abs/1811.04968} {arXiv:1811.04968} \BibitemShut {NoStop}%
\bibitem [{\citenamefont {Developers}(2024)}]{developers_cirq_2024}%
  \BibitemOpen
  \bibfield  {author} {\bibinfo {author} {\bibfnamefont {C.}~\bibnamefont {Developers}},\ }\href {https://doi.org/10.5281/zenodo.11398048} {\bibinfo {title} {Cirq}} (\bibinfo {year} {2024})\BibitemShut {NoStop}%
\bibitem [{\citenamefont {Chatterjee}\ \emph {et~al.}(2023)\citenamefont {Chatterjee}, \citenamefont {Das}, \citenamefont {Bala}, \citenamefont {Saha}, \citenamefont {Chattopadhyay},\ and\ \citenamefont {Chakrabarti}}]{chatterjee_qudiet:_2023}%
  \BibitemOpen
  \bibfield  {author} {\bibinfo {author} {\bibfnamefont {T.}~\bibnamefont {Chatterjee}}, \bibinfo {author} {\bibfnamefont {A.}~\bibnamefont {Das}}, \bibinfo {author} {\bibfnamefont {S.~K.}\ \bibnamefont {Bala}}, \bibinfo {author} {\bibfnamefont {A.}~\bibnamefont {Saha}}, \bibinfo {author} {\bibfnamefont {A.}~\bibnamefont {Chattopadhyay}},\ and\ \bibinfo {author} {\bibfnamefont {A.}~\bibnamefont {Chakrabarti}},\ }\bibfield  {title} {\bibinfo {title} {{QuDiet}: {A} classical simulation platform for qubit‐qudit hybrid quantum systems},\ }\href {https://doi.org/10.1049/qtc2.12058} {\bibfield  {journal} {\bibinfo  {journal} {IET Quantum Communication}\ }\textbf {\bibinfo {volume} {4}},\ \bibinfo {pages} {167} (\bibinfo {year} {2023})}\BibitemShut {NoStop}%
\bibitem [{\citenamefont {Vincent}\ \emph {et~al.}(2022)\citenamefont {Vincent}, \citenamefont {O'Riordan}, \citenamefont {Andrenkov}, \citenamefont {Brown}, \citenamefont {Killoran}, \citenamefont {Qi},\ and\ \citenamefont {Dhand}}]{vincent_jet:_2022}%
  \BibitemOpen
  \bibfield  {author} {\bibinfo {author} {\bibfnamefont {T.}~\bibnamefont {Vincent}}, \bibinfo {author} {\bibfnamefont {L.~J.}\ \bibnamefont {O'Riordan}}, \bibinfo {author} {\bibfnamefont {M.}~\bibnamefont {Andrenkov}}, \bibinfo {author} {\bibfnamefont {J.}~\bibnamefont {Brown}}, \bibinfo {author} {\bibfnamefont {N.}~\bibnamefont {Killoran}}, \bibinfo {author} {\bibfnamefont {H.}~\bibnamefont {Qi}},\ and\ \bibinfo {author} {\bibfnamefont {I.}~\bibnamefont {Dhand}},\ }\bibfield  {title} {\bibinfo {title} {Jet: {Fast} quantum circuit simulations with parallel task-based tensor-network contraction},\ }\href {https://doi.org/10.22331/q-2022-05-09-709} {\bibfield  {journal} {\bibinfo  {journal} {Quantum}\ }\textbf {\bibinfo {volume} {6}},\ \bibinfo {pages} {709} (\bibinfo {year} {2022})}\BibitemShut {NoStop}%
\bibitem [{\citenamefont {Kreplin}\ \emph {et~al.}(2024)\citenamefont {Kreplin}, \citenamefont {Willmann}, \citenamefont {Schnabel}, \citenamefont {Rapp}, \citenamefont {Hagelüken},\ and\ \citenamefont {Roth}}]{kreplin2024squlearn}%
  \BibitemOpen
  \bibfield  {author} {\bibinfo {author} {\bibfnamefont {D.~A.}\ \bibnamefont {Kreplin}}, \bibinfo {author} {\bibfnamefont {M.}~\bibnamefont {Willmann}}, \bibinfo {author} {\bibfnamefont {J.}~\bibnamefont {Schnabel}}, \bibinfo {author} {\bibfnamefont {F.}~\bibnamefont {Rapp}}, \bibinfo {author} {\bibfnamefont {M.}~\bibnamefont {Hagelüken}},\ and\ \bibinfo {author} {\bibfnamefont {M.}~\bibnamefont {Roth}},\ }\href@noop {} {\bibinfo {title} {squlearn -- a python library for quantum machine learning}} (\bibinfo {year} {2024}),\ \Eprint {https://arxiv.org/abs/2311.08990} {arXiv:2311.08990} \BibitemShut {NoStop}%
\bibitem [{\citenamefont {Wang}\ \emph {et~al.}(2022)\citenamefont {Wang}, \citenamefont {Ding}, \citenamefont {Gu}, \citenamefont {Li}, \citenamefont {Lin}, \citenamefont {Pan}, \citenamefont {Chong},\ and\ \citenamefont {Han}}]{torchquantum}%
  \BibitemOpen
  \bibfield  {author} {\bibinfo {author} {\bibfnamefont {H.}~\bibnamefont {Wang}}, \bibinfo {author} {\bibfnamefont {Y.}~\bibnamefont {Ding}}, \bibinfo {author} {\bibfnamefont {J.}~\bibnamefont {Gu}}, \bibinfo {author} {\bibfnamefont {Z.}~\bibnamefont {Li}}, \bibinfo {author} {\bibfnamefont {Y.}~\bibnamefont {Lin}}, \bibinfo {author} {\bibfnamefont {D.~Z.}\ \bibnamefont {Pan}}, \bibinfo {author} {\bibfnamefont {F.~T.}\ \bibnamefont {Chong}},\ and\ \bibinfo {author} {\bibfnamefont {S.}~\bibnamefont {Han}},\ }\bibfield  {title} {\bibinfo {title} {Quantumnas: Noise-adaptive search for robust quantum circuits},\ }in\ \href@noop {} {\emph {\bibinfo {booktitle} {The 28th IEEE International Symposium on High-Performance Computer Architecture (HPCA-28)}}}\ (\bibinfo {year} {2022})\BibitemShut {NoStop}%
\bibitem [{\citenamefont {Mato}\ \emph {et~al.}(2024)\citenamefont {Mato}, \citenamefont {Ringbauer}, \citenamefont {Burgholzer},\ and\ \citenamefont {Wille}}]{mqtqudits}%
  \BibitemOpen
  \bibfield  {author} {\bibinfo {author} {\bibfnamefont {K.}~\bibnamefont {Mato}}, \bibinfo {author} {\bibfnamefont {M.}~\bibnamefont {Ringbauer}}, \bibinfo {author} {\bibfnamefont {L.}~\bibnamefont {Burgholzer}},\ and\ \bibinfo {author} {\bibfnamefont {R.}~\bibnamefont {Wille}},\ }\href {https://arxiv.org/abs/2410.02854} {\bibinfo {title} {Mqt qudits: A software framework for mixed-dimensional quantum computing}} (\bibinfo {year} {2024}),\ \Eprint {https://arxiv.org/abs/2410.02854} {arXiv:2410.02854} \BibitemShut {NoStop}%
\bibitem [{\citenamefont {Chollet}\ \emph {et~al.}(2015)\citenamefont {Chollet} \emph {et~al.}}]{keras}%
  \BibitemOpen
  \bibfield  {author} {\bibinfo {author} {\bibfnamefont {F.}~\bibnamefont {Chollet}} \emph {et~al.},\ }\href@noop {} {\bibinfo {title} {Keras}},\ \bibinfo {howpublished} {\url{https://keras.io}} (\bibinfo {year} {2015})\BibitemShut {NoStop}%
\bibitem [{\citenamefont {Preskill}(2018)}]{Preskill_2018}%
  \BibitemOpen
  \bibfield  {author} {\bibinfo {author} {\bibfnamefont {J.}~\bibnamefont {Preskill}},\ }\bibfield  {title} {\bibinfo {title} {Quantum computing in the nisq era and beyond},\ }\href {https://doi.org/10.22331/q-2018-08-06-79} {\bibfield  {journal} {\bibinfo  {journal} {Quantum}\ }\textbf {\bibinfo {volume} {2}},\ \bibinfo {pages} {79} (\bibinfo {year} {2018})}\BibitemShut {NoStop}%
\bibitem [{\citenamefont {Callison}\ and\ \citenamefont {Chancellor}(2022)}]{hybrid1}%
  \BibitemOpen
  \bibfield  {author} {\bibinfo {author} {\bibfnamefont {A.}~\bibnamefont {Callison}}\ and\ \bibinfo {author} {\bibfnamefont {N.}~\bibnamefont {Chancellor}},\ }\bibfield  {title} {\bibinfo {title} {Hybrid quantum-classical algorithms in the noisy intermediate-scale quantum era and beyond},\ }\href {https://doi.org/10.1103/PhysRevA.106.010101} {\bibfield  {journal} {\bibinfo  {journal} {Phys. Rev. A}\ }\textbf {\bibinfo {volume} {106}},\ \bibinfo {pages} {010101} (\bibinfo {year} {2022})}\BibitemShut {NoStop}%
\bibitem [{\citenamefont {Chen}\ \emph {et~al.}(2021)\citenamefont {Chen}, \citenamefont {Zhao}, \citenamefont {Song}, \citenamefont {Zhao}, \citenamefont {Wang},\ and\ \citenamefont {Wang}}]{hybrid2}%
  \BibitemOpen
  \bibfield  {author} {\bibinfo {author} {\bibfnamefont {R.-Y.-L.}\ \bibnamefont {Chen}}, \bibinfo {author} {\bibfnamefont {B.-C.}\ \bibnamefont {Zhao}}, \bibinfo {author} {\bibfnamefont {Z.-X.}\ \bibnamefont {Song}}, \bibinfo {author} {\bibfnamefont {X.-Q.}\ \bibnamefont {Zhao}}, \bibinfo {author} {\bibfnamefont {K.}~\bibnamefont {Wang}},\ and\ \bibinfo {author} {\bibfnamefont {X.}~\bibnamefont {Wang}},\ }\bibfield  {title} {\bibinfo {title} {Hybrid quantum-classical algorithms: Foundation, design and applications},\ }\href {https://doi.org/10.7498/aps.70.20210985} {\bibfield  {journal} {\bibinfo  {journal} {Acta Physica Sinica}\ }\textbf {\bibinfo {volume} {70}},\ \bibinfo {pages} {210302} (\bibinfo {year} {2021})}\BibitemShut {NoStop}%
\bibitem [{\citenamefont {Deutsch}\ and\ \citenamefont {Jozsa}(1992)}]{Deutsch1992RapidSO}%
  \BibitemOpen
  \bibfield  {author} {\bibinfo {author} {\bibfnamefont {D.}~\bibnamefont {Deutsch}}\ and\ \bibinfo {author} {\bibfnamefont {R.}~\bibnamefont {Jozsa}},\ }\bibfield  {title} {\bibinfo {title} {Rapid solution of problems by quantum computation},\ }\href {https://api.semanticscholar.org/CorpusID:121702767} {\bibfield  {journal} {\bibinfo  {journal} {Proceedings of the Royal Society of London. Series A: Mathematical and Physical Sciences}\ }\textbf {\bibinfo {volume} {439}},\ \bibinfo {pages} {553 } (\bibinfo {year} {1992})}\BibitemShut {NoStop}%
\bibitem [{\citenamefont {Mogos}(2008)}]{deutsch_qudit1}%
  \BibitemOpen
  \bibfield  {author} {\bibinfo {author} {\bibfnamefont {G.}~\bibnamefont {Mogos}},\ }\bibfield  {title} {\bibinfo {title} {The deutsch-josza algorithm for n-qudits},\ }\href@noop {} {\bibfield  {journal} {\bibinfo  {journal} {International Journal of Computers, Communications and Control (IJCCC)}\ } (\bibinfo {year} {2008})}\BibitemShut {NoStop}%
\bibitem [{\citenamefont {Marttala}(2007)}]{deutsch_qudit2}%
  \BibitemOpen
  \bibfield  {author} {\bibinfo {author} {\bibfnamefont {P.}~\bibnamefont {Marttala}},\ }\emph {\bibinfo {title} {An extension of the Deutsch-Jozsa algorithm to arbitrary qudits}},\ \href {http://hdl.handle.net/10388/etd-07262007-105041} {\bibinfo {type} {Master of science (m.sc.) thesis}},\ \bibinfo  {school} {University of Saskatchewan} (\bibinfo {year} {2007})\BibitemShut {NoStop}%
\bibitem [{\citenamefont {Grover}(1996)}]{grover}%
  \BibitemOpen
  \bibfield  {author} {\bibinfo {author} {\bibfnamefont {L.~K.}\ \bibnamefont {Grover}},\ }\bibfield  {title} {\bibinfo {title} {A fast quantum mechanical algorithm for database search},\ }in\ \href {https://doi.org/10.1145/237814.237866} {\emph {\bibinfo {booktitle} {Proceedings of the Twenty-Eighth Annual ACM Symposium on Theory of Computing}}},\ \bibinfo {series and number} {STOC '96}\ (\bibinfo  {publisher} {Association for Computing Machinery},\ \bibinfo {address} {New York, NY, USA},\ \bibinfo {year} {1996})\ p.\ \bibinfo {pages} {212–219}\BibitemShut {NoStop}%
\bibitem [{\citenamefont {Nikolaeva}\ \emph {et~al.}(2023)\citenamefont {Nikolaeva}, \citenamefont {Kiktenko},\ and\ \citenamefont {Fedorov}}]{grover_qudit1}%
  \BibitemOpen
  \bibfield  {author} {\bibinfo {author} {\bibfnamefont {A.~S.}\ \bibnamefont {Nikolaeva}}, \bibinfo {author} {\bibfnamefont {E.~O.}\ \bibnamefont {Kiktenko}},\ and\ \bibinfo {author} {\bibfnamefont {A.~K.}\ \bibnamefont {Fedorov}},\ }\bibfield  {title} {\bibinfo {title} {Generalized toffoli gate decomposition using ququints: Towards realizing grover's algorithm with qudits},\ }\bibfield  {journal} {\bibinfo  {journal} {Entropy}\ }\textbf {\bibinfo {volume} {25}},\ \href {https://doi.org/10.3390/e25020387} {10.3390/e25020387} (\bibinfo {year} {2023})\BibitemShut {NoStop}%
\bibitem [{\citenamefont {Saha}\ \emph {et~al.}(2022)\citenamefont {Saha}, \citenamefont {Majumdar}, \citenamefont {Saha}, \citenamefont {Chakrabarti},\ and\ \citenamefont {Sur-Kolay}}]{grover_qudit2}%
  \BibitemOpen
  \bibfield  {author} {\bibinfo {author} {\bibfnamefont {A.}~\bibnamefont {Saha}}, \bibinfo {author} {\bibfnamefont {R.}~\bibnamefont {Majumdar}}, \bibinfo {author} {\bibfnamefont {D.}~\bibnamefont {Saha}}, \bibinfo {author} {\bibfnamefont {A.}~\bibnamefont {Chakrabarti}},\ and\ \bibinfo {author} {\bibfnamefont {S.}~\bibnamefont {Sur-Kolay}},\ }\bibfield  {title} {\bibinfo {title} {Asymptotically improved circuit for a d-ary grover's algorithm with advanced decomposition of the n-qudit toffoli gate},\ }\bibfield  {journal} {\bibinfo  {journal} {Physical Review A}\ }\textbf {\bibinfo {volume} {105}},\ \href {https://doi.org/10.1103/physreva.105.062453} {10.1103/physreva.105.062453} (\bibinfo {year} {2022})\BibitemShut {NoStop}%
\bibitem [{\citenamefont {Cerezo}\ \emph {et~al.}(2021{\natexlab{b}})\citenamefont {Cerezo}, \citenamefont {Arrasmith}, \citenamefont {Babbush}, \citenamefont {Benjamin}, \citenamefont {Endo}, \citenamefont {Fujii}, \citenamefont {McClean}, \citenamefont {Mitarai}, \citenamefont {Yuan}, \citenamefont {Cincio},\ and\ \citenamefont {Coles}}]{Cerezo_2021}%
  \BibitemOpen
  \bibfield  {author} {\bibinfo {author} {\bibfnamefont {M.}~\bibnamefont {Cerezo}}, \bibinfo {author} {\bibfnamefont {A.}~\bibnamefont {Arrasmith}}, \bibinfo {author} {\bibfnamefont {R.}~\bibnamefont {Babbush}}, \bibinfo {author} {\bibfnamefont {S.~C.}\ \bibnamefont {Benjamin}}, \bibinfo {author} {\bibfnamefont {S.}~\bibnamefont {Endo}}, \bibinfo {author} {\bibfnamefont {K.}~\bibnamefont {Fujii}}, \bibinfo {author} {\bibfnamefont {J.~R.}\ \bibnamefont {McClean}}, \bibinfo {author} {\bibfnamefont {K.}~\bibnamefont {Mitarai}}, \bibinfo {author} {\bibfnamefont {X.}~\bibnamefont {Yuan}}, \bibinfo {author} {\bibfnamefont {L.}~\bibnamefont {Cincio}},\ and\ \bibinfo {author} {\bibfnamefont {P.~J.}\ \bibnamefont {Coles}},\ }\bibfield  {title} {\bibinfo {title} {Variational quantum algorithms},\ }\href {https://doi.org/10.1038/s42254-021-00348-9} {\bibfield  {journal} {\bibinfo  {journal} {Nature Reviews Physics}\ }\textbf {\bibinfo {volume} {3}},\ \bibinfo {pages} {625–644} (\bibinfo {year}
  {2021}{\natexlab{b}})}\BibitemShut {NoStop}%
\bibitem [{\citenamefont {Fisher}(1988)}]{misc_iris_53}%
  \BibitemOpen
  \bibfield  {author} {\bibinfo {author} {\bibfnamefont {R.~A.}\ \bibnamefont {Fisher}},\ }\href@noop {} {\bibinfo {title} {{Iris}}},\ \bibinfo {howpublished} {UCI Machine Learning Repository} (\bibinfo {year} {1988}),\ \bibinfo {note} {{DOI}: https://doi.org/10.24432/C56C76}\BibitemShut {NoStop}%
\bibitem [{\citenamefont {LeCun}\ \emph {et~al.}(2010)\citenamefont {LeCun}, \citenamefont {Cortes},\ and\ \citenamefont {Burges}}]{lecun2010mnist}%
  \BibitemOpen
  \bibfield  {author} {\bibinfo {author} {\bibfnamefont {Y.}~\bibnamefont {LeCun}}, \bibinfo {author} {\bibfnamefont {C.}~\bibnamefont {Cortes}},\ and\ \bibinfo {author} {\bibfnamefont {C.}~\bibnamefont {Burges}},\ }\bibfield  {title} {\bibinfo {title} {Mnist handwritten digit database},\ }\href@noop {} {\bibfield  {journal} {\bibinfo  {journal} {ATT Labs [Online]. Available: http://yann.lecun.com/exdb/mnist}\ }\textbf {\bibinfo {volume} {2}} (\bibinfo {year} {2010})}\BibitemShut {NoStop}%
\bibitem [{\citenamefont {Pavlidis}\ and\ \citenamefont {Floratos}(2021)}]{Pavlidis_2021}%
  \BibitemOpen
  \bibfield  {author} {\bibinfo {author} {\bibfnamefont {A.}~\bibnamefont {Pavlidis}}\ and\ \bibinfo {author} {\bibfnamefont {E.}~\bibnamefont {Floratos}},\ }\bibfield  {title} {\bibinfo {title} {Quantum-fourier-transform-based quantum arithmetic with qudits},\ }\bibfield  {journal} {\bibinfo  {journal} {Physical Review A}\ }\textbf {\bibinfo {volume} {103}},\ \href {https://doi.org/10.1103/physreva.103.032417} {10.1103/physreva.103.032417} (\bibinfo {year} {2021})\BibitemShut {NoStop}%
\end{thebibliography}%


\appendix{}

\onecolumngrid

\section{UML diagram of the QuForge library}
\label{appendix:1}

In this appendix, we provide a detailed UML class diagram of the QuForge library to complement the discussion in the main text. The diagram illustrates the core data structures (\texttt{State}), the circuit construction and execution engine (\texttt{Circuit} and \texttt{Gate} hierarchy), the optimization interface (\texttt{CircuitWrapper} and \texttt{optim}), and the measurement utilities. Standard UML notation is used to indicate class attributes, operations, and inter-class relationships (composition, dependency, and utility usage), offering a comprehensive overview of the object-oriented architecture of the library.

\begin{figure}[H]
    \centering
    \includegraphics[width=1.0\linewidth]{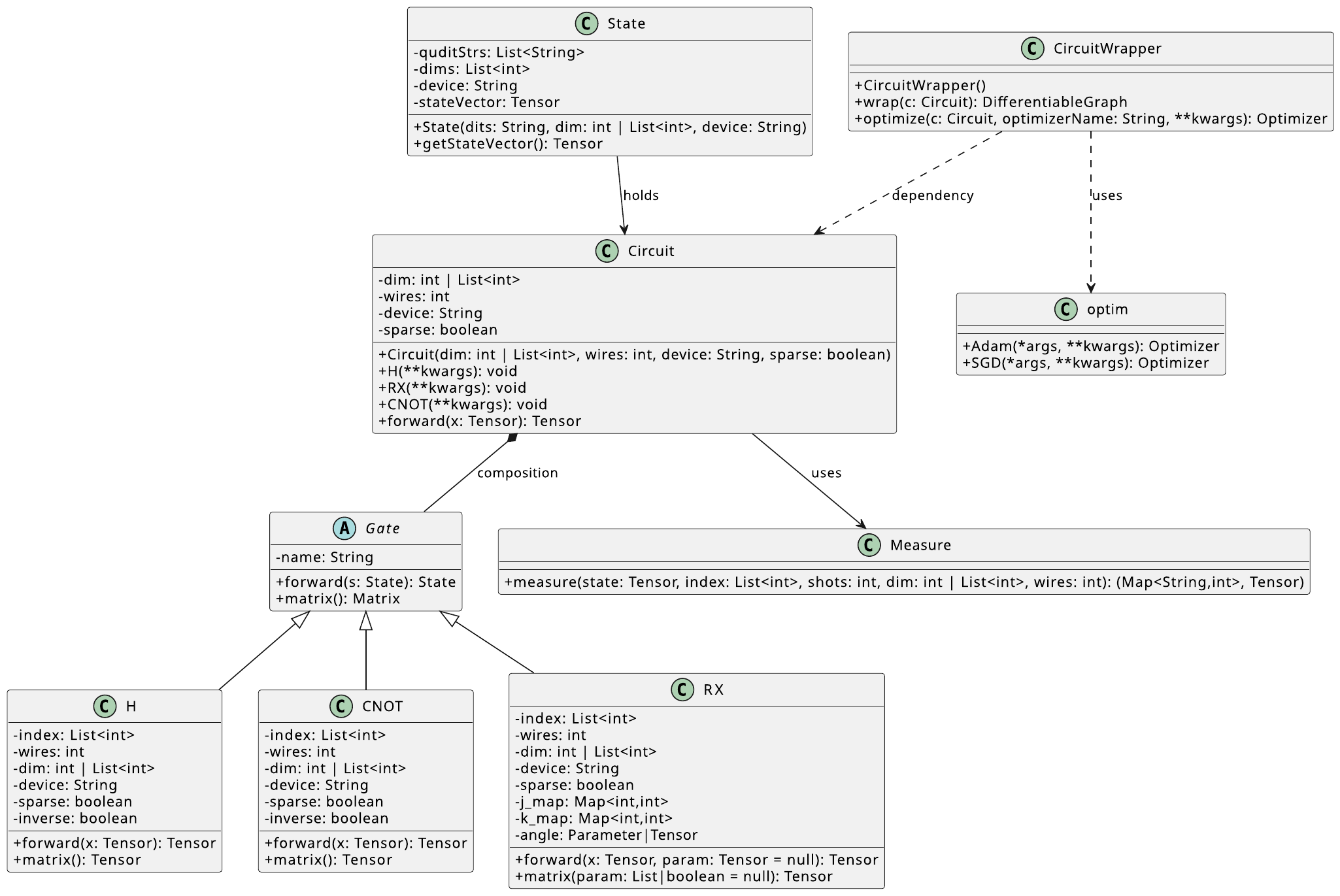}
    \caption{UML class diagram of the QuForge library. Core components include the \texttt{State} class for qudit state construction, the \texttt{Circuit} module (with gate‐adding and forward execution methods), the \texttt{Gate} hierarchy (showing a few gates such as \texttt{H}, \texttt{CNOT}, and \texttt{RX}), the \texttt{CircuitWrapper} (with optimization integration), and the standalone measurement utility.}
    \label{fig:diagram}
\end{figure}

\section{Matrix form of the quantum gates}
\label{appendix:2}

This appendix provides a comprehensive and detailed construction of the quantum gates. Each gate is described, with explicit definitions of its respective operations. Furthermore, when feasible, the derivation of the sparse matrix representation for each gate is presented. Table \ref{table:2} summarizes the main gates available.

\begin{table}[h!]
\setlength{\tabcolsep}{0pt}
\begin{center}
\begin{tabular}{ | c | c |} 
  \hline
  Single qudit gates & Two-qudit gates \\
  \hline
  \hline
  NOT, Eq. (\ref{gate:NOT}) & CNOT, Eq. (\ref{gate:cnot}) \\
  \hline
  Phase-shift, Eq. (\ref{gate:phase}) &  SWAP, Eq. ( \ref{gate:swap}) \\
  \hline
  Fourier, Eq. (\ref{gate:fourier}) & Controlled Rotation, Eq. (\ref{gate:cr}) \\
  \hline
\end{tabular}
\caption{Overview of the primary single-qudit and two-qudit gates available in the QuForge library.}
\label{table:2}
\end{center}
\end{table}


\subsection{Generalized NOT gate} \label{gate:NOT}

The \texttt{X} gate, also known as the $NOT$ gate, is fundamental for qubits. It performs a bit-flip operation, transforming the state of a qubit from $|0\rangle$ to $|1\rangle$ and vice versa. When dealing with qudits, the analogous operation is more intricate. In this context, the \texttt{X} gate performs a modulus sum on the qudit's value. Typically, this summand, denoted as $s$, is set to 1, although other values can be used depending on the desired operation. Consequently, this gate induces a cyclic permutation of the qudit states. For example, in a three-level system, the application of the \texttt{X} gate results in the following transitions: $|0\rangle \rightarrow |1\rangle \rightarrow |2\rangle \rightarrow |0\rangle$. This gate has the following representation:
\begin{equation}
    X^D_{s} |x\rangle = |x+s \bmod D\rangle.
\end{equation}

\subsection{Generalized phase-shift gate}\label{gate:phase}


Phase-shift gates are essential components in quantum computing, particularly for manipulating the phase of qudits in $D$-level quantum systems. These gates adjust the amplitude of the qudit states by applying a phase factor. Let $\omega=e^{2\pi i/D}$, where $D$ denotes the dimension of the qudit. The phase-shift gate operates as follows: it leaves the state $|0\rangle$ unchanged, i.e., $|0\rangle \rightarrow |0\rangle$, while transforming the state $|1\rangle$ to $\omega|1\rangle$, the state $|2\rangle$ to $\omega^2|2\rangle$, and so forth. This progression continues cyclically through all $D$ states. The general transformation for a state $|k\rangle$ can be expressed as $|k\rangle\rightarrow\omega^k|k\rangle$.

The corresponding unitary matrix representation of this phase-shift gate, denoted as $P$, is a diagonal matrix with the elements $\{1,\omega,\omega^2,...,\omega^{D-1}\}$ along its diagonal. Formally, the matrix is given by:
\begin{equation}
    P = 
    \begin{bmatrix}
        1 & 0 & 0 & ... & 0\\
        0 & \omega & 0 & ... & 0\\
        0 & 0 & \omega^2 & ... & 0\\
        \vdots & \vdots & \vdots & \vdots & \vdots \\
        0 & 0 & 0 & ... & w^{D-1}
    \end{bmatrix}.
\end{equation}

\subsection{Fourier gate}\label{gate:fourier}

The Fourier gate, also called the generalized Hadamard gate, plays an important role in quantum computing by enabling the creation of superposition states. The nature of the superposition generated by this gate depends intricately on the initial state of the qudit to which it is applied. When applied to the state $|0\rangle$, the Fourier gate produces an equal superposition of all $D-1$ possible states, each with identical amplitude. For other initial states, the resulting amplitudes vary according to the specific starting state of the qudit, thereby introducing a rich structure of quantum superpositions.

The matrix representation of the Fourier gate, denoted as \texttt{H}, is derived from the principles of the discrete Fourier transform. This matrix can be expressed as:
\begin{equation}
    H =
    \frac{1}{\sqrt{D}}
    \begin{bmatrix}
        1 & 1 & 1 & ... & 1\\
        1 & \omega & \omega^2 & ... & \omega^{D-1}\\
        1 & \omega^2 & \omega^4 & ... & \omega^{2(D-1)}\\
        \vdots & \vdots & \vdots & \vdots & \vdots \\
        1 & \omega^{D-1} & \omega^{2(D-1)} & ... & \omega^{(D-1)^2}
    \end{bmatrix}.
\end{equation}
The factor $\frac{1}{\sqrt{D}}$ ensures the normalization of the transformation, maintaining the unitary property of the gate.

\subsection{Generalized Gell-Mann matrices}

The generalized Gell-Mann matrices, also known as generalized Pauli matrices, are important in the context of rotation gates in quantum computing. These matrices extend the concept of the Pauli matrices to higher-dimensional systems and play an important role in defining rotations and other transformations in $D$-level quantum systems. They come in three distinct forms: symmetric, anti-symmetric, and diagonal.


The generalized Gell-Mann matrices are defined as follows:
\begin{enumerate}
\item Symmetric form:
\begin{align}
    S_x^{jk} &= |j\rangle\langle k| + |k\rangle \langle j|.
\end{align}
This matrix, with $1\le j< k\le D$,
represents a symmetric combination of the basis states $|j\rangle$ and $|k\rangle$, facilitating rotations that symmetrically affect these states.

\item Anti-symmetric form:
\begin{align}
    S_y^{jk} &= -i|j\rangle\langle k| + i|k\rangle \langle j|.
\end{align}
Here, for $1\le j< k\le D$,
we have an imaginary component,
useful for generating rotations that are skew-symmetric, affecting the states $|j\rangle$ and $|k\rangle$ with opposite phases.

\item Diagonal form:
\begin{align}
    S_z^{j} &= \sqrt{\frac{2}{j(j+1)}}\sum_{k=1}^j (-j)^{\delta(k,j)}|k\rangle \langle k|
\end{align}
The diagonal form of the matrix is defined for each $j=1,\cdots,D-1$
and involves a summation over the basis states up to $j$. The Kronecker delta $\delta(k,j)$ ensures that the contribution is weighted appropriately, introducing a factor that depends on the index $j$.
\end{enumerate}

\subsection{Generalized rotation gate}\label{gate:rotation} 
Rotation gates are among the most crucial elements in quantum machine learning, owing to their ability to introduce tunable parameters that can be optimized to achieve specific outcomes in quantum circuits. These gates are characterized by an angle parameter $\theta$, which can be adjusted during the training of a quantum algorithm to fine-tune the performance of the quantum model.
The implementation of rotation gates relies fundamentally on the generalized Gell-Mann matrices. These matrices serve as the generators for the rotations, controlling the axis and nature of the rotation within the $D$-dimensional quantum state space. The general form of a rotation gate is given by:
\begin{equation}
    R_{\alpha} = e^{-i\theta S_{\alpha}/2},
\end{equation}
where $S_\alpha$ represents the generalized Gell-Mann matrix, and $\alpha$ denotes the matrix type (symmetric, anti-symmetric, or diagonal). 

In practical applications, the parameter $\theta$ is typically adjusted using optimization algorithms to minimize a cost function, enabling the quantum circuit to learn and perform specific tasks effectively. This process is analogous to adjusting weights in classical machine learning models. The flexibility offered by rotation gates makes them fundamental in constructing variational quantum circuits, which are central to many quantum machine learning frameworks.

For the sparse form of the rotation gate, we start from
\begin{equation}
R_\alpha^{j,k}(\theta) = e^{-i\theta S_\alpha^{j,k}/2},
\end{equation}
with $\alpha = x, y, z$. To better understand the construction and action of these rotation gates, we can explore specific cases and then generalize the pattern.

\textbf{Case $D=2$.} For $D=2$, we have a single possibility for each direction:
\begin{align}
S_x^{1,2} & = |1\rangle\langle 2| + |2\rangle\langle 1| = \begin{bmatrix}0 & 1 \\ 1 & 0\end{bmatrix}, \\ 
S_y^{1,2} & = -i|1\rangle\langle 2| + i|2\rangle\langle 1| = \begin{bmatrix}0 & -i \\ i & 0\end{bmatrix}, \\
S_z^{1} & = \sum_{k=1}^{2}\sqrt{1}(-1)^{\delta_{k,2}}|k\rangle\langle k| \nonumber \\
& = |1\rangle\langle 1| - |2\rangle\langle 2| = \begin{bmatrix}1 & 0 \\ 0 & -1\end{bmatrix}.
\end{align}
Here, the states $|1\rangle$ and $|2\rangle$ form the computational basis for the qubit system.
The corresponding rotation matrices for $D=2$ are:
\begin{align}
& R_x^{1,2}(\theta) = e^{-i\theta S_x^{1,2}/2} = e^{(-i\theta/2)(|+\rangle\langle +| - |-\rangle\langle -|)} \nonumber \nonumber \\ 
& = \begin{bmatrix}\cos(\theta/2) & -i\sin(\theta/2) \\ -i\sin(\theta/2) & \cos(\theta/2)\end{bmatrix}, \\
& R_y^{1,2}(\theta) = e^{-i\theta S_y^{1,2}/2} = e^{(-i\theta/2)(|\oplus\rangle\langle \oplus| - |\ominus\rangle\langle \ominus|)} \nonumber \\ 
& = \begin{bmatrix}\cos(\theta/2) & -\sin(\theta/2) \\ \sin(\theta/2) & \cos(\theta/2)\end{bmatrix}, \\
& R_z^{1}(\theta) = e^{-i\theta S_z^{1}/2} = e^{(-i\theta/2)(|1\rangle\langle 1| - |2\rangle\langle 2|)} \nonumber \\ 
& = \begin{bmatrix}e^{-i\theta/2} & 0 \\ 0 & e^{i\theta/2}\end{bmatrix},
\end{align}
with $|\pm\rangle = (|1\rangle\pm|2\rangle)/\sqrt{2}$, $|\oplus\rangle = (|1\rangle + i |2\rangle)/\sqrt{2}$, and $|\ominus\rangle = (|1\rangle -i|2\rangle)/\sqrt{2}$.

\textbf{Case for $D=3$.} For $D=3$, we consider a few examples to identify the general pattern:
\begin{align}
& S_x^{1,2} = |1\rangle\langle 2| + |2\rangle\langle 1| = \begin{bmatrix}0 & 1 & 0 \\ 1 & 0 & 0 \\ 0 & 0 & 0\end{bmatrix}.
\end{align}
Thus
\begin{align}
R_x^{1,2}(\theta) & = e^{-i\theta S_x^{1,2}/2} = e^{(-i\theta/2)(|+\rangle\langle +| - |-\rangle\langle -|)} \nonumber \\ 
& = \begin{bmatrix}\cos(\theta/2) & -i\sin(\theta/2) & 0 \\ -i\sin(\theta/2) & \cos(\theta/2) & 0 \\ 0 & 0 & 1\end{bmatrix}.
\end{align}

\textbf{Case for $D=4$.} For $D=4$, let us consider
\begin{align}
& S_x^{1,2} = |1\rangle\langle 2| + |2\rangle\langle 1| = \begin{bmatrix}0 & 1 & 0 & 0 \\ 1 & 0 & 0 & 0 \\ 0 & 0 & 0 & 0 \\ 0 & 0 & 0 & 0\end{bmatrix}.
\end{align}
Thus
\begin{align}
R_x^{1,2}(\theta) & = e^{-i\theta S_x^{1,2}/2} = e^{(-i\theta/2)(|+\rangle\langle +| - |-\rangle\langle -|)} \nonumber \\ 
& = \begin{bmatrix}\cos(\theta/2) & -i\sin(\theta/2) & 0 & 0 \\ -i\sin(\theta/2) & \cos(\theta/2) & 0 & 0 \\ 0 & 0 & 1 & 0 \\ 0 & 0 & 0 & 1\end{bmatrix}.
\end{align}


In the general case, this pattern repeats for any dimension. Specifically, the two-dimensional rotation is applied in the subspace with indices \(j, k\), while the identity operation is applied to the rest of the vector space. Then, the non-zero elements of \(R_x^{j,k}\) are:
\begin{align}
& \langle j|R_x^{j,k}(\theta)|j\rangle = \langle k|R_x^{j,k}(\theta)|k\rangle = \cos(\theta/2), \\ 
& \langle j|R_x^{j,k}(\theta)|k\rangle = \langle k|R_x^{j,k}(\theta)|j\rangle = -i\sin(\theta/2),\\
& \langle l|R_x^{j,k}(\theta)|l\rangle = 1, \quad l \ne j \ \text{and} \ l \ne k.
\end{align}
And the non-zero elements of \(R_y^{j,k}\) are:
\begin{align}
& \langle j|R_y^{j,k}(\theta)|j\rangle = \langle k|R_y^{j,k}(\theta)|k\rangle = \cos(\theta/2), \\ 
& \langle j|R_y^{j,k}(\theta)|k\rangle = -\langle k|R_y^{j,k}(\theta)|j\rangle = -\sin(\theta/2),\\
& \langle l|R_y^{j,k}(\theta)|l\rangle = 1, \quad l \ne j \ \text{and} \ l \ne k,
\end{align}
while the non-zero elements of \(R_z^{j}\) are:
\begin{align}
& \langle k|R_z^{j}(\theta)|k\rangle = \exp\left(-i\frac{\theta}{2}\sqrt{\frac{2}{j(j+1)}}(-j)^{\delta_{k,j+1}}\right),
\end{align}
for $k=1,\cdots,j-1$ and
\begin{equation}
\langle k|R_z^{j}(\theta)|k\rangle = 1
\end{equation}
for $k=j,\cdots,d.$

It is worth mentioning that we worked with the generalized Gell-Mann matrices using the computational basis $\{|j\rangle\}_{j=1}^D$, for convenience. But when programming the analytical results, we apply $\{|j\rangle\}_{j=0}^{D-1}.$

\subsection{Generalized CNOT gate}\label{gate:cnot}

The generalized \texttt{CNOT} gate, also known as the \texttt{CX} gate, is a fundamental two-qudit gate in quantum computing, essential for creating entanglement between two qudits. This gate operates with two components: the control qudit and the target qudit. The role of the \texttt{CNOT} gate is to preserve the state of the control qudit while modifying the state of the target qudit based on the value of the control qudit by applying an \texttt{X} gate.

The functionality of the generalized \texttt{CNOT} gate can be described as follows: if the control qudit is in the state $|x\rangle$, the state of the target qudit is changed to $|x+y \bmod D\rangle$, effectively performing an addition modulo $D$, as defined with the $X$ gate.
Mathematically, the action of the generalized \texttt{CNOT} gate is represented by:
\begin{equation}
    CNOT_{x \rightarrow y} |x\rangle |y\rangle = 
    \begin{cases}
      |x\rangle |x+a \mbox{ mod } D\rangle, \mbox{ if } x = d-1\\
      |x\rangle |y\rangle, \mbox{ otherwise }
    \end{cases}\,.
\end{equation}

A computational basis state of N qudits can be represented as: 
\begin{align}
& |j_0\rangle \otimes |j_1\rangle \otimes \cdots \otimes |j_{N-2}\rangle \otimes |j_{N-1}\rangle \\
& \equiv |j_0 j_1 \cdots j_{N-2} j_{N-1}\rangle \equiv |j\rangle, 
\end{align}
where the local state index $j_s$ ranges from $0$ to $D-1$, and the qudit index $s$ ranges from $0$ to $N-1$. The correspondence to the global state index, known as the decimal representation, is calculated as follows: 
\begin{equation}
 j = \sum_{k=0}^{N-1} j_k d^{N-1-k} .
\end{equation}

To determine the non-zero elements of the \texttt{CNOT} gate matrix for N qudits, we must consider the control qudit (denoted by index $c$) and the target qudit (denoted by index $t$). Depending on the relative positions of $c$ and $t$ (i.e., whether $c < t$ or $c > t$, the non-zero elements of the \texttt{CNOT} gate matrix are obtained as follows:
$\bullet$ For $c < t$:
\begin{equation} 
\langle \cdots, k_{c-1}, k_c, k_{c+1}, \cdots, k_{t-1}, k_t \oplus k_c, k_{t+1}, \cdots | C_x | \cdots, k_{c-1}, k_c, k_{c+1}, \cdots, k_{t-1}, k_t, k_{t+1}, \cdots \rangle = 1.
\end{equation}
$\bullet$ For $c > t$:
\begin{equation}
\langle \cdots, k_{t-1}, k_t \oplus k_c, k_{t+1}, \cdots, k_{c-1}, k_c, k_{c+1}, \cdots | C_x | \cdots, k_{t-1}, k_t, k_{t+1}, \cdots, k_{c-1}, k_c, k_{c+1}, \cdots \rangle = 1
\end{equation}

In both cases, $\oplus$ denotes addition modulo $D$. 
To generate the non-zero elements of the \texttt{CNOT} gate matrix, we start by generating all possible sequences of local basis states for N qudits, each of dimension $D$. These sequences correspond to the computational basis states $|j_0,j_1,...,j_{N-2}, j_{N-1}\rangle$. The decimal representation of these sequences gives us the matrix indices for the right-hand side of the matrix element.

Next, we apply the transformation $j_t\rightarrow j_t \oplus j_c$ to obtain the matrix index for the left-hand side corresponding to the target qudit. This process ensures that we correctly identify the positions of the non-zero elements in the \texttt{CNOT} gate matrix, reflecting the gate's action of conditionally flipping the target qudit's state based on the control qudit's state.

\subsection{Generalized SWAP Gate}
\label{gate:swap}

The \texttt{SWAP} gate is a two-qudit gate, designed to exchange the states of two qudits. This gate performs a simple operation, swapping the state of the first qudit with that of the second qudit. The action of the \texttt{SWAP} gate can be mathematically described as follows:
\begin{equation}
SWAP|x\rangle \otimes |y\rangle = |y\rangle \otimes |x\rangle
\end{equation}
Here, $|x\rangle$ and $|y\rangle$ represent the states of the two qudits. When the \texttt{SWAP} gate is applied to the tensor product of these states, it interchanges them, resulting in the state $|y\rangle \otimes |x\rangle$.

For two qudits, the non-zero elements of the \texttt{SWAP} gate matrix are determined by examining how the gate acts on the basis states. The matrix element $\langle l,m|SWAP|p,q\rangle$ represents the overlap between the basis states $|l,m\rangle$ and $|p,q\rangle$. When the \texttt{SWAP} gate is applied, it exchanges the states of the two qudits, meaning that $SWAP|p,q\rangle$ results in $|q,p\rangle$. Therefore, the non-zero elements are found where the original states $|p,q\rangle$ and the swapped states $|l,m\rangle$ match, which occurs if and only if $l=q$ and $m=p$. This condition is mathematically expressed using the Kronecker delta function: $\delta_{l,p}\delta_{m,q}$.

For an $N$-qudit system, applying the \texttt{SWAP} gate to qudits $c$ and $t$ involves a similar procedure but within a larger state space. The matrix element $\langle j|SWAP_{c\leftrightarrow t}|k\rangle$ reflects the action of the \texttt{SWAP} gate on multi-qudit states. Here, the \texttt{SWAP} gate exchanges the states of the $c$-th and $t$-th qudits while leaving the other qudits' states unchanged.

To find the non-zero elements, consider the basis state $|\cdots k_{c-1}k_c k_{c+1}\cdots k_{t-1}k_t k_{t+1}\cdots\rangle$. After applying the \texttt{SWAP} gate, this state transforms to $|\cdots k_{c-1}k_t k_{c+1}\cdots k_{t-1}k_c k_{t+1}\cdots\rangle$. Therefore, the matrix element $\langle j|SWAP_{c\leftrightarrow t}|k\rangle$ is non-zero when the sequence of indices $j$ matches this swapped configuration.

Explicitly, this means that the conditions for non-zero elements are:
\begin{align}
& \langle \cdots j_{c-1}j_c j_{c+1}\cdots j_{t-1}j_t j_{t+1}\cdots|\cdots k_{c-1}k_t k_{c+1}\cdots k_{t-1}k_c k_{t+1}\cdots\rangle \\
& =
\cdots\delta_{j_{c-1},k_{c-1}}\delta_{j_c,k_t}\delta_{j_{c+1},k_{c+1}}\cdots\delta_{j_{t-1},k_{t-1}}\delta_{j_t,k_c}\delta_{j_{t+1},k_{t+1}}\cdots
\end{align}

Thus, to determine the non-zero elements of the \texttt{SWAP} matrix, we systematically identify pairs of states $j$ and $k$ that satisfy these delta conditions, effectively swapping the $c$-th and $t$-th components while leaving the others unchanged.

\subsection{Controlled Rotation Gate}
\label{gate:cr}

The controlled rotation gate operates similar to the generalized \texttt{CNOT} gate. It preserves the state of the control qudit while applying a rotation gate to the target qudit. In this context, the controlled rotation gate can be viewed as a more general form of the \texttt{CNOT} gate. Precisely, when the controlled rotation gate uses the symmetric form of the generalized Gell-Mann matrices and an angle $\theta=\pi$, it replicates the behavior of the \texttt{CNOT} gate.

To formalize this gate, it is essential to note that previous definitions, such as the one in reference \cite{Pavlidis_2021}, describe a non-differentiable controlled gate that does not align with the controlled rotation gate for qubits. Hence, we propose a new differentiable version of the controlled rotation gate that maintains equivalence for qubits while extending its applicability to qudits.

The differentiable controlled rotation gate is defined as:
\begin{equation}
    C^{c\rightarrow t}_{R^{jk}_{\alpha}(\theta)} = \sum_{m=0}^{D-1} |m\rangle \langle m| \otimes R^{jk}_{\alpha}(m\theta),
\end{equation}
with $R^{jk}_{\alpha}(\theta)=e^{-i\theta S^{jk}_\alpha/2}.$
In this expression, the gate operates on the control qudit ($c$) and the target qudit ($t$). The operator $|m\rangle \langle m|$ acts as a projector on the control qudit, ensuring that the rotation is conditioned on the state of the control qudit. The term $e^{-im\theta S_\alpha/2}$ represents the rotation applied to the target qudit, with $S_\alpha$ being the generalized Gell-Mann matrix. This construction ensures that the controlled rotation gate is smooth and differentiable, which is important for optimization tasks in quantum machine learning. By enabling precise control over rotations, this gate facilitates the implementation of complex quantum algorithms that leverage the rich structure of qudit systems.

Regarding the sparse implementation of the controlled rotation gate, let us define e.g. $\mathbf{j} = \cdots,k_{c-1},k_c,k_{c+1},\cdots,k_{t-1},j,k_{t+1},\cdots$ if $t>c$ and $\mathbf{j}=\cdots,k_{t-1},j,k_{t+1},\cdots,k_{c-1},k_c,k_{c+1},\cdots$ if $t<c$, where $k_s = 0,\cdots,d-1$ and $s=1,\cdots,N$ with $N$ being the number of qudits and $d$ is the dimension of each qudit. For the $x$-type rotations, the non-null elements are
\begin{align}
& \langle \mathbf{j}|C^{c\rightarrow t}_{R^{jk}_{x}(\theta)}|\mathbf{j}\rangle = \langle \mathbf{k}|C^{c\rightarrow t}_{R^{jk}_{x}(\theta)}|\mathbf{k}\rangle = \cos(k_c \theta/2), \\
& \langle \mathbf{j}|C^{c\rightarrow t}_{R^{jk}_{x}(\theta)}|\mathbf{k}\rangle = \langle \mathbf{k}|C^{c\rightarrow t}_{R^{jk}_{x}(\theta)}|\mathbf{j}\rangle = -i\sin(k_c \theta/2), \\
& \langle \mathbf{l}|C^{c\rightarrow t}_{R^{jk}_{x}(\theta)}|\mathbf{l}\rangle = 1;\  l=0,\cdots,d-1;\ l\ne j,l\ne k.
\end{align}
In the case of the $y$-type rotations, the non-null elements are given by
\begin{align}
& \langle \mathbf{j}|C^{c\rightarrow t}_{R^{jk}_{y}(\theta)}|\mathbf{j}\rangle = \langle \mathbf{k}|C^{c\rightarrow t}_{R^{jk}_{x}(\theta)}|\mathbf{k}\rangle = \cos(k_c \theta/2), \\
& \langle \mathbf{j}|C^{c\rightarrow t}_{R^{jk}_{y}(\theta)}|\mathbf{k}\rangle = \langle \mathbf{k}|C^{c\rightarrow t}_{R^{jk}_{x}(\theta)}|\mathbf{j}\rangle = -\sin(k_c \theta/2), \\
& \langle \mathbf{l}|C^{c\rightarrow t}_{R^{jk}_{y}(\theta)}|\mathbf{l}\rangle = 1;\  l=0,\cdots,d-1;\ l\ne j,l\ne k.
\end{align}
For the $z$-type rotations, the non-null elements are
\begin{equation}
\langle\mathbf{k}|C_{R_z^{j}(\theta)}^{c\rightarrow t}|\mathbf{k}\rangle =  \exp\Big(-i\frac{k_c \theta}{2}\sqrt{\frac{2}{j(j+1)}}(-j)^{\delta_{k,j+1}}\Big)
\end{equation}
for $k=1,\cdots,j+1$
and
\begin{equation}
\langle\mathbf{k}|C_{R_z^{j}(\theta)}^{c\rightarrow t}|\mathbf{k}\rangle =  1
\end{equation}
for $k=j+2,\cdots,d.$




\end{document}